\documentclass{article}
\usepackage{graphicx}                    
\usepackage{rotating}
\usepackage{color}                       

\chardef\us=`\_

\newcommand{\lya}{Ly$\rm \alpha$}

\newcommand{\FLone}{SOL2010--02--08}
\newcommand{\FLtwo}{SOL2016--04--18}
\newcommand{\FLthree}{SOL2023--05--09}

\usepackage{natbib}

\usepackage{authblk}  
\usepackage{orcidlink}  

\title{On the Instrumental Discrepancies in Lyman-alpha Observations of Solar Flares}

\author[1]{Harry J. Greatorex\orcidlink{0000-0002-5302-0887}\thanks{Corresponding author: hgreatorex01@qub.ac.uk}}
\author[1]{Ryan O. Milligan\orcidlink{0000-0001-5031-1892}}
\author[2]{Ingolf E. Dammasch\orcidlink{0000-0002-5302-0887}}

\affil[1]{\small{Astrophysics Research Centre, School of Mathematics and Physics, Queen's University Belfast, Belfast, United Kingdom, BT7 1NN}}
\affil[2]{Solar Influences Data Analysis Center, Royal Observatory of Belgium, Circular Avenue 3, 1180 Uccle, Brussels, Belgium}

\date{}  

\begin{document}
\maketitle

\markboth{Greatorex et al.}{Solar Physics Example Article}

\begin{abstract}
Despite the energetic significance of Lyman-alpha (\lya; 1216\AA) emission from solar flares, regular observations of flare related \lya\ have been relatively scarce until recently. Advances in instrumental capabilities and a shift in focus over previous Solar Cycles mean it is now routinely possible to take regular co-observations of \lya\ emission in solar flares. Thus, it is valuable to examine how the instruments selected for flare observations may influence the conclusions drawn from the analysis of their unique measurements. Here, we examine three M-class flares each observed in \lya\ by GOES-14/EUVS-E, GOES-15/EUVS-E, or GOES-16/EXIS-EUVS-B, and at least one other instrument from PROBA2/LYRA, MAVEN/EUVM, ASO-S/LST-SDI, and SDO/EVE-MEGS-P. For each flare, the relative and excess flux, contrast, total energy, and timings of the \lya\ emission were compared between instruments. It was found that while the discrepancies in measurements of the relative flux between instruments may be considered minimal, the calculated contrasts, excess fluxes, and energetics may differ significantly - in some cases up to a factor of five. This may have a notable impact on multi-instrument investigations of the variable \lya\ emission in solar flares and estimates of the contribution of \lya\ to the radiated energy budget of the chromosphere. The findings presented in this study will act as a guide for the interpretation of observations of flare-related \lya\ from upcoming instruments during future Solar Cycles and inform conclusions drawn from multi-instrument studies.
\end{abstract}



%

\section{Introduction} \label{sec:intro}

The Lyman-alpha line of neutral hydrogen (\lya; 1216\AA) is the most intense emission line in the quiescent solar spectrum \citep{Curdt2001LyaBrightest}. In spite of this, observations of flare-related \lya\ emission have historically been relatively scarce. Among the earliest measurements of flare-related \lya\ emission were the spectroscopic observations of \cite{Skumanich1978FlareLya} using the \textit{Laboratoire de Physique Stellaire et Planetaire} (LPSP; \citealt{Bonnet1978LPSP_OSO8}) onboard \textit{Eighth Orbiting Solar Observatory} (OSO-8), and of \cite{CanfieldvanHoosier1980} using the NRL Spectrograph \citep{Bartoe1977NRLSpec} within the \textit{Apollo Telescope Mount} onboard \textit{Skylab}. However, due to the individual instrumental capabilities required to observe such an energetic emission line on flare timescales, it has only been possible in the last decade to conduct statistical studies using calibrated photometric measurements of flare-related \lya.

\cite{Milligan2020MXFlares} conducted a large-scale statistical study of \lya\ flares observed by the \textit{Geostationary Operational Environmental Satellites} (GOES) during Solar Cycle 24. From their analysis of $>$500 M- and X-class flares, it was found that both the energy and contrast in flare-related \lya\ emission tend to scale with Soft X-Ray (SXR) magnitude. Furthermore, it was found that 95\% of the flares sampled had an associated \lya\ contrast of $<$10\%. However, unique cases could reach contrasts of up to $\sim$30\%. Comparatively, observations of flare-related \lya\ from the \textit{Project for On-Board Autonomy} (PROBA2; \citealt{Santandrea2013Proba2}) have consistently found contrasts of  $\rm<1\%$ (\citealt{Kretzshmar2013LyraFLare, Raulin2013LyaIonosphere}). One explanation for these notably low contrasts may be significant detector degradation and the presence of contaminants in the PROBA2 signal. Currently, the extent to which these discrepancies may exist in co-observations between individual instruments is not well documented.

\lya\ emission is optically thick and therefore observations in \lya\ are subject to effects of Centre-to-Limb Variation (CLV), whereby the position of a flare on the solar disk (and the subsequent column depth along the observing line of sight) can impact the measured \lya\ irradiance (for further discussion see \citealt{Woods1995CLV, Woods2006TSIVariation, Milligan2021B&CClassFlares}). The extent to which CLV may impact \lya\ observations was examined by \cite{Milligan2020MXFlares} using combined stereoscopic observations of an X1.1 flare from GOES-15 and the \textit{Mars Atmosphere and Volatile Evolution} (MAVEN; \citealt{Eparvier2015MavenEUVM}) in orbit around Mars. For MAVEN, the flare was located close to disk-centre, whereas for GOES the flare appeared at the solar limb. Subsequently, a $\sim$45\% increase in ﬂare-related \lya\ excess was observed at Mars relative to Earth, thus demonstrating the magnitude of CLV on observations of optically-thick emission.

Temporally, flare-related \lya\ emission tends to peak in conjunction with nonthermal Hard X-Ray (HXR) emission and therefore the derivative of the SXRs in alignment with the Neupert Effect \citep{Neupert1968neuperteffect}; examples of the nonthermal origin of \lya\ emission are demonstrated in \cite{Nusinov2006LyaNTCorr}, \cite{deCosta2009NTLyaCorr}, \cite{Milligan2017NTLyaCorr}, \cite{Dominique2018NTLyaCorr}, \cite{Li2022C14SolO}, \cite{Tian2023ntlya}, and \cite{Greatorex2023EquivMagFlares}. In some instances, secondary \lya\ emission peaks occur during the decay-phase, which is speculated to originate from cooling loop plasma rather than the flare footpoint regions \citep{Kretzshmar2013LyraFLare}, or from filament eruptions \citep{Wauters2022M67Lya}. However, systematic effects have also been shown to account for apparent violations of expected Neupertian behaviour. \cite{Milligan2016SDO_EVE_MEGSP} found flare-related \lya\ emission observed by the \textit{Extreme-ultraviolet Variability Experiment} (EVE; \citealt{Woods2012EVE}) onboard SDO to peak within the gradual phase co-temporally with the associated Soft X-ray (SXR) emission. This was later attributed to issues with the Kalman filters used during data processing (D. Woodraska - Private Communication)  

Recent instrumental advancements mean that it is now possible to image solar flares in \lya\ on sufficient timescales to observe the dynamic spatial development of flare footpoints, loops, and ribbons. \cite{Li2022C14SolO} studied a C1.4 flare using observations from the \lya\ channel of \textit{High Resolution Imager} within the \textit{Extreme Ultraviolet Imager} onboard \textit{Solar Orbiter} (SO/EUI-HRI$_{Ly\alpha}$; \citealt{Muller2020SolarOrbiter, Rochus2020EUI}\footnote{SO was omitted from this study due to difficulty in obtaining sufficient calibration data.}). In their study, the authors found a co-spatial relationship between \lya\ emission and nonthermal HXR sources, similarity in spatiotemporal behaviour between \lya\ and He~{\sc{ii}} (304\AA) emission observed by the \textit{Extreme Ultraviolet Imager} (EUVI; \citealt{Howard2008EUVI}) onboard the \textit{Solar Terrestrial Relations Observatory Ahead} (STEREO-A; \citealt{Kaiser2008Stereo}) and the \textit{Atmospheric Imaging Assembly} onboard the \textit{Solar Dynamics Observatory} (SDO/AIA; \citealt{Pesnell2012SDOMain, Boerner2012AIA}), and \lya\ brightenings in both the rise and decay phase of the flare. Furthermore, the \textit{Solar Disk Imager} within the \textit{Lyman-alpha Solar Telescope} onboard the recently launched \textit{Advanced Space-based Solar Observatory} (ASO-S/SDI-LST; \citealt{Li2019LST, Feng2019LST, Gan2023ASOS, Chen2024LST}) has been used to examine the presence of long-period QPPs in an X6.4 solar flare, co-temporal with the associated HXR emission \citep{Dong2024LSTQPP}. Similar behaviour was presented by \cite{Milligan2017NTLyaCorr} in a multi-instrument study using observations from GOES, SDO, and the \textit{Reuven Ramaty High Energy Solar Spectroscopic Imager} (RHESSI; \citealt{Lin2002RHESSI}).

Multi-instrument observations are crucial for examining fundamental open questions surrounding the origin of flare-related emission, constraining the radiated energy budget of the chromosphere, and the temporal evolution of flare-related emission during different phases of solar flares. Using joint observations from SDO, GOES, PROBA2, and RHESSI, \cite{Wauters2022M67Lya} examined an M6.7 flare with an additional peak in \lya\ emission during the decay phase of the flare that had no temporal correlation to the HXR or SXR emission. Using images in 1600\AA\ from SDO/AIA, the source of the emission was later attributed to a failed filament eruption in proximity to the flare origin. Moreover, using combined observations from RHESSI and GOES, both \cite{Milligan2014EUVEnergy} and \cite{Greatorex2023EquivMagFlares} were able to examine the contribution of \lya\ to the radiated energy budget of the chromosphere during solar flares. These studies found for both an X- and M-class flare, \lya\ may account for up to $\sim$8\% of the energy deposited into the chromosphere by nonthermal electrons. 

When conducting precise analysis of solar flares, it is currently unclear to what extent the instruments chosen for observations may impact the conclusions drawn from their observations. This fact becomes more pertinent with the consideration of newly-launched and upcoming missions such as ASO-S, GOES-R, and Solar-C \citep{Watanabe2014SolarC}. With the wealth of available data from current missions, it is possible to thoroughly examine \lya\ emission during solar flares. This paper aims to assess the level of agreement in \lya\ observations of solar flares between different missions, focussing on observable metrics that may be used to infer the underlying mechanisms driving the emission, such as the relative and excess fluxes, the flare contrasts, energetics, and timings. This analysis should act as a guide for future studies using next generation instruments. 

The instruments examined within this study are summarised in Section~\ref{sec:instmeth}. A discussion of the unique instrumental observing capabilities including the spectral response functions and bandpasses is presented in Section~\ref{sec:bpsrf}. The flare sample used for the individual case studies, as well as the calibration, standardisation, and analysis techniques are presented in Section~\ref{sec:cal&meth}. Section~\ref{sec:results} details the results of each case study. Finally, a discussion of the impact of this study and future \lya\ observation capabilities is presented in Section~\ref{sec:Discussion}.

\section{Instrumentation} \label{sec:instmeth}

This study focuses on the examination of observations of flare-related \lya\ emission from multiple missions that have served the solar community over Solar Cycle 24 and 25. The following section presents a brief summary of each instrument included in this study and details the individual calibration processes carried out for each instrument where relevant. Table \ref{table:instrument_summary} contains a technical summary of each instrument.

\subsection{GOES-14/EUVS--E and GOES-15/EUVS--E} \label{subsec:g14andg15}

GOES-14 and GOES-15 both featured an \textit{X-Ray Sensor} (XRS; \citealt{Hanser1996GOES_XRS, Chamberlin2009XRS}) and \textit{Extreme Ultraviolet Sensor} (EUVS; \citealt{Viereck2007EUVS, Evans2010EUVS}), observing disk-integrated SXR and Extreme Ultraviolet (EUV) emission at a near 100\% duty cycle. EUVS consisted of five channels (A--E) covering the 50--170\AA, 240--340\AA, 200--620\AA, 200--800\AA, and 1180--1250\AA\ wavelength ranges. The E-channel (EUVS--E) was a dedicated broadband channel centred around the \lya\ line at 1216\AA, observing the full solar disk at 10.24s cadence. These GOES satellites operated in a geostationary orbit at an altitude of approximately 36 000~km. Thus, observations from GOES are subject to geocoronal absorption that are more pronounced around the equinoxes. In this instance, EUV emission is absorbed by hydrogen in the geocorona, leading to a reduction in the observed emission. As a result, it is important to consider \lya\ observations where the effect of the geocorona is minimal and/or can be accounted for. The GOES-14/EUVS-E and GOES-15/EUVS-E observations suffer from degradation over time, which is compensated for by using daily averages from the \textit{Solar Stellar Irradiance Comparison Experiment} onboard the \textit{Solar Radiation and Climate Experiment} (SORCE/SOLSTICE; \citealt{McClintock2005SOLSTICE}) to scale the daily averages from GOES. Thus, while the degradation and absolute values are scaled to SORCE/SOLSTICE, the variability is determined by the GOES instrument. For this study, the degradation corrected 1nm band (121-122nm) data were used.

\subsection{GOES-16/EXIS--EUVS--B} \label{subsec:g16}

GOES-16 operates with an incorporated set of SXR and EUV sensors as part of a combined \textit{Extreme Ultraviolet and X-ray Irradiance Sensors} (EXIS) suite. The EUVS for EXIS is comprised of two channels (A and B) measuring the spectral irradiance of lines within the 250--310\AA\ and 1170--1410\AA\ wavelength ranges, respectively, and a third channel (Channel-C) taking relative measurements of the Mg~{\sc{ii}} core/wing ratio between 2750--2850\AA. The EUVS--B channel is comprised of a photodiode array with diode clusters allocated to cover significant solar emission lines such as C~{\sc{iii}} (1175\AA), H~{\sc{i}} \lya\ (1216\AA), C~{\sc{ii}} (1335\AA), and S~{\sc{iv}}/O~{\sc{iv}} (1405\AA), each with an approximate width of 6\AA. The final irradiance data from EUVS-B is a summation of the fluxes over the photodiodes for each separate cluster. At the time of writing, these full-disk L2 irradiances from EXIS/EUVS--B are available at 60s cadence, future products following additional processing are expected to have cadences of 1s. Similar to red GOES-14/15, GOES-16 operates in a geostationary orbit with an altitude of approximately 36 000~km (see \citealt{Eparvier2009EUVS} for a detailed discussion of EUVS for GOES-16). For this study, Version 1.0.5 of the science-quality solar line data from GOES-16/EXIS were used; at the time of writing this is the most up-to-date publicly available version of the GOES-16 data. 

Several caveats need to be considered for GOES-16/EXIS-EUVS-B data. Firstly, the presence of multi-hour post-eclipse thermal dips in the spectral lines due to incompletely corrected temperature impacts may affect the EUVS observations, although it is simple to exclude these data as required. The magnitude of this in the \lya\ line is difficult to quantify due to the additional effect of the geocorona. Furthermore, an annual cycle oscillation artifact with a magnitude of 1.3\% of the measured flux impacts the irradiances of the \lya\ channel of EUVS. It is unclear what impact these may have on the observations presented in this study.
  
\subsection{PROBA2/LYRA} \label{subsec:lyra}

The \textit{Large Yield Radiometer} (LYRA; \citealt{Hochedez2006PROBA2Lyra, Dominique2013LYRAInstr}) onboard PROBA2 is a tri-unit broadband radiometer with four distinct channels observing in the SXR to mid-ultraviolet regime. The \lya\ filters from Units~1 and 2 (MSM Diamond Detector) and Unit~3 (AXUV Si Detector) of LYRA cover the 1200--1230\AA\ wavelength range, taking full-disk irradiance measurements at 0.05s cadence. The nominal unit of LYRA has suffered significant degradation and no longer observes \lya\ emission, due to changes in the \lya\ channel bandpass. The degradation of the LYRA instrument has been attributed to the deposition of carbon and silicon on the detectors, creating a contaminant layer that is more opaque to longer wavelengths relative to short resulting in different signal losses depending on wavelength (\citealt{BenMoussa2015LyraDeg, Wauters2022M67Lya}). The data collected during special observing campaigns using the backup unit are still sufficient to observe flare related \lya\ emission (\textit{M. Dominique. 2023 - Private Communication}). PROBA2 has a Sun-synchronous orbit at 720~km altitude placing it in Low-Earth Orbit. 

\subsection{MAVEN/EUVM} \label{subsec:maven}

The \textit{Extreme Ultraviolet Monitor} (EUVM) onboard the MAVEN satellite is a Sun-facing instrument consisting of three channels taking disk-integrated broadband irradiances in the SXR and EUV regimes. The \lya\ channel (Channel C) of EUVM covers the 1170--1250\AA\ wavelength range, measuring irradiance at 1s cadence. Nominal EUVM observations are taken in the operations case with the Sun within science FOV and aperture mechanism “open”, which produces valid solar irradiance measurements. MAVEN operates in a highly elliptical orbit around Mars, with an apoapsis of $\sim$6000~km and periapsis of $\sim$150~km.

\subsection{SDO/EVE-MEGS-P} \label{subsec:sdo}

SDO/EVE features a set of \textit{Multiple EUV Grating Spectrographs} (MEGS) comprised of three main components. The grazing-incidence spectrograph (MEGS-A) and normal-incidence spectrograph (MEGS-B) sample the 50-370\AA\ and 350-1050\AA\ wavelength range at 1\AA\ resolution, respectively\footnote{Due to an instrumentation failure on 26~May~2014, MEGS-A is no longer operational.}. The broadband photodiode (MEGS-P) used with the first grating of MEGS-B is centred on the \lya\ line. SDO operates in a 28$^{\circ}$ inclined geosynchronous orbit
at an altitude of 36 000~km. At the time of writing, the degradation of MEGS-B means that MEGS-P operates on a flare trigger system, with a 60s cadence (previously MEGS-P has operated with 10s cadence).

\subsection{ASO-S/LST-SDI} \label{subsec:asos}

The ASO-S mission is the first space-based Chinese mission dedicated to solar observation. The LST is one of the instruments onboard ASO-S. The SDI within LST images the full solar disk at 1216\AA\ with a spatial resolution of $9.5''$ and a routine exposure time of 13.5s\footnote{This has increased to 16.5s due to degradation of the instrument.}. ASO-S operates in a Sun-synchronous orbit with an altitude of 720~km and an inclination of 98$^{\circ}$ relative to Earth's equator. The nominal cadence for routine SDI observations is between 4--40s; SDI images are also available in 60s intervals due to the ``\textit{image download interval}'' set in response to telemetry limitations.

\begin{table}[!ht]
\centering
\caption{Summary of the observing instruments from each mission used in this work. *Cadences are taken as the cadence of the L2 calibrated data available from the dedicated mission repositories for each observation. **Average between two peaks in GOES-15/EUVS-E response function.}\label{table:instrument_summary}
\hspace*{-2.5cm}
\begin{tabular}{lllllll}
\hline
\textbf{Mission} & \textbf{Instrument} & \textbf{Orbit}  & \textbf{Observation} & \textbf{Cadence} & \textbf{Bandpass} & \textbf{Response Peak} \\
\vspace{-0.4cm}
\\
\vspace{0cm}
& & & \textbf{Type} & \textbf{(s)*} & \textbf{FWHM (\AA)} & \textbf{$\lambda$ (\AA)} \\
\hline
\vspace{0.2cm}
GOES-14       & EUVS--E       & Geostationary   & Photometry & 10.24        & 131.1                              & 1214                            \\
\vspace{0.2cm}
GOES-15       & EUVS--E    & Geostationary   & Photometry & 10.24        & 108.7                              & 1227**                \\
\vspace{0.2cm}
GOES-16       & EXIS--EUVS--B  & Geostationary   & Photometry  & 60.0           & -                                  & -                               \\
\vspace{0.2cm}
PROBA2        & LYRA    & Sun-synchronous & Photometry & 0.05          & 104.5                              & 1200                            \\
\vspace{0.2cm}
MAVEN         & EUVM   & Areocentric     & Photometry  & 1.0            & 64.4                               & 1200                            \\
\vspace{0.2cm}
SDO           & EVE-MEGS-P   & Geosynchronous  & Photometry   & 60.0           & 91.6                               & 1205                            \\
\vspace{0.2cm}
ASO-S         & LST-SDI     & Sun-synchronous & Imaging     & 60.0           & 92.3                               & 1216                            \\
\vspace{0.2cm}
              &                 &                    &                      &              &                                    &                                
\end{tabular}
\end{table}

\section{Spectral Response Functions and Bandpasses} \label{sec:bpsrf}

\begin{figure}[!ht]    
\centerline{\includegraphics[width=1\textwidth]{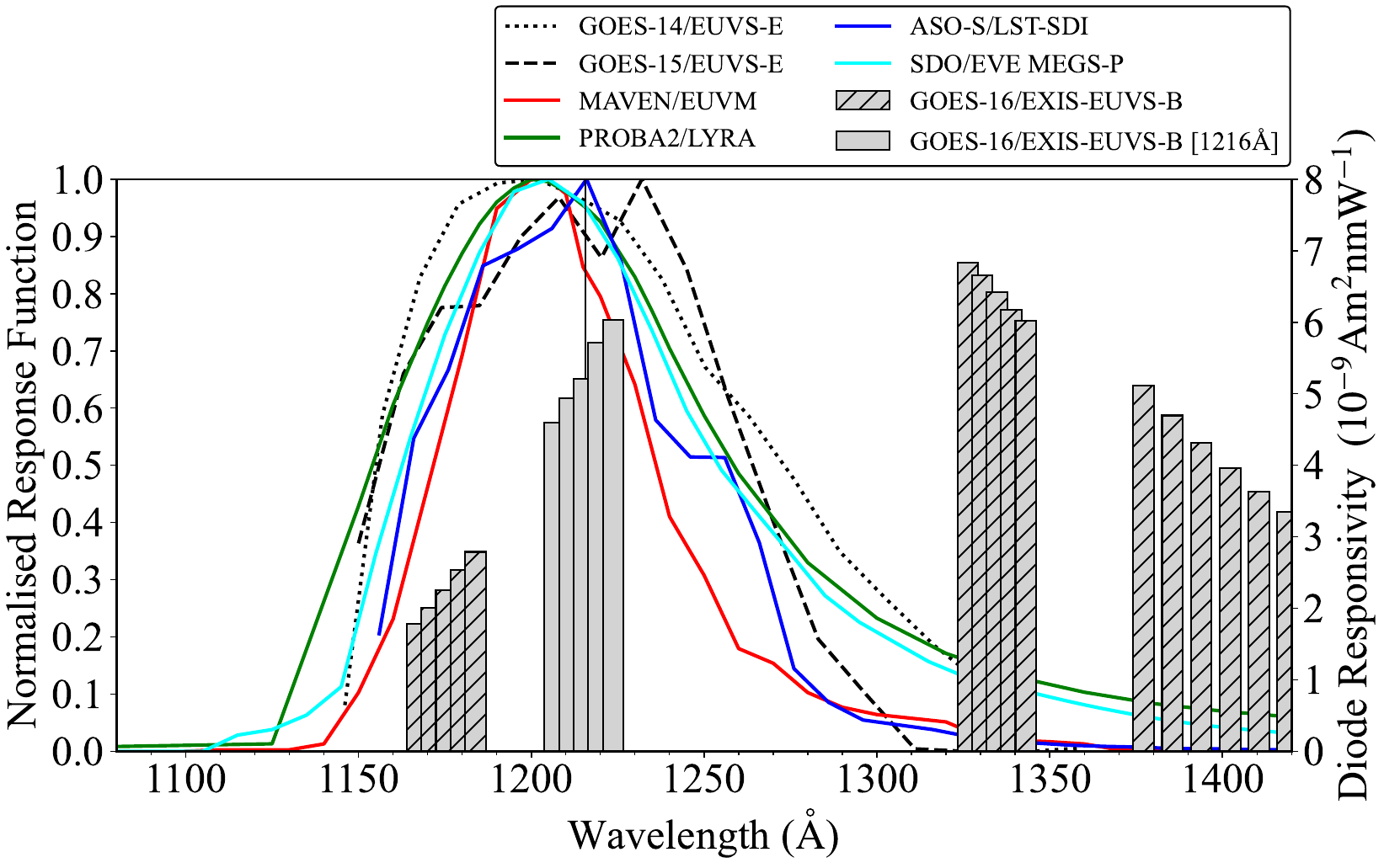}}
\caption{Normalised spectral responses for the instruments examined in this work. The \lya\ line core at 1216\AA\ is denoted by the vertical solid line. The GOES-16 diodes are denoted by grey bars, where the width of the bar is approximately 6\AA. Unhatched grey bars denote the cluster about the \lya\ line.}
\label{figure_response}
\end{figure}

The spectral response of a given instrument quantifies the effectiveness of an instrument in transforming incident flux into a measurable output for a given emission spectrum. Figure~\ref{figure_response} shows the normalised spectral response for each broadband instrument described in Section~\ref{sec:instmeth}. Also presented are the diode responsivities of GOES-16/EXIS--EUVS--B, where each grey bar denotes a unique photodiode. Generally, the normalised spectral responses for the broadband \lya\ instruments resemble a Gaussian profile with the central peak residing within the 1200-1220\AA\ wavelength range. It appears that several of the spectral responses are centred slightly blueward of the \lya\ core at 1216\AA\ (denoted by the vertical solid black line in Figure~\ref{figure_response}). The GOES-15/EUVS--E spectral response appears to be double peaked, with the larger peak appearing around 1245\AA\ and the average of the two peaks being $\sim$1227\AA. The responsivities for GOES-16/EXIS--EUVS--B show the distinct clusters centred about their unique corresponding emission lines, with cluster~B positioned about the \lya\ core and the responsivity apparently increasing monotonically with wavelength. It should be noted that the responses are normalised for comparison and do not necessarily give an indication of the true instrumental responsivity as they are dimensionless. The true responsivity may change due to degradation over the instrument lifetime. The wavelength at the peak of the spectral response for each instrument is summarised in Table \ref{table:instrument_summary}. 

\begin{figure}[!ht]    
\centerline{\includegraphics[width=0.9\textwidth]{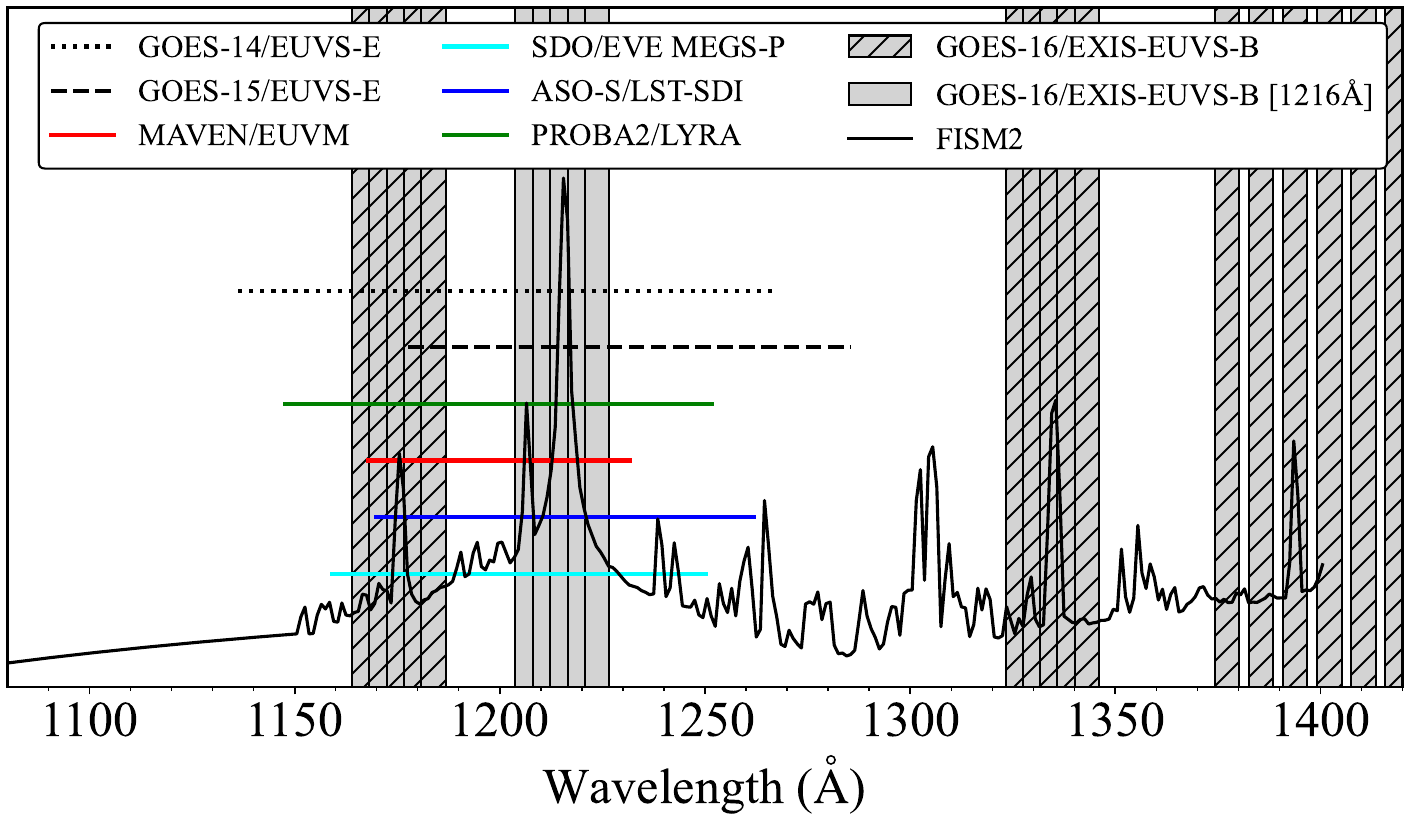}}
\caption{FWHM of the spectral responses presented in Figure~\ref{figure_response}. The horizontal lines denote the widths of the FWHM bandpasses. The GOES-16 diodes are denoted by grey bars, where the width of each bar is approximately 6\AA. Unhatched grey bars denote the cluster about the \lya\ line. A QS EUV spectrum from FISM2 is overplot in the solid, thin black line.} 
\label{figure_bandpass}
\end{figure}

\begin{figure}[!ht]    
\centerline{\includegraphics[width=1\textwidth]{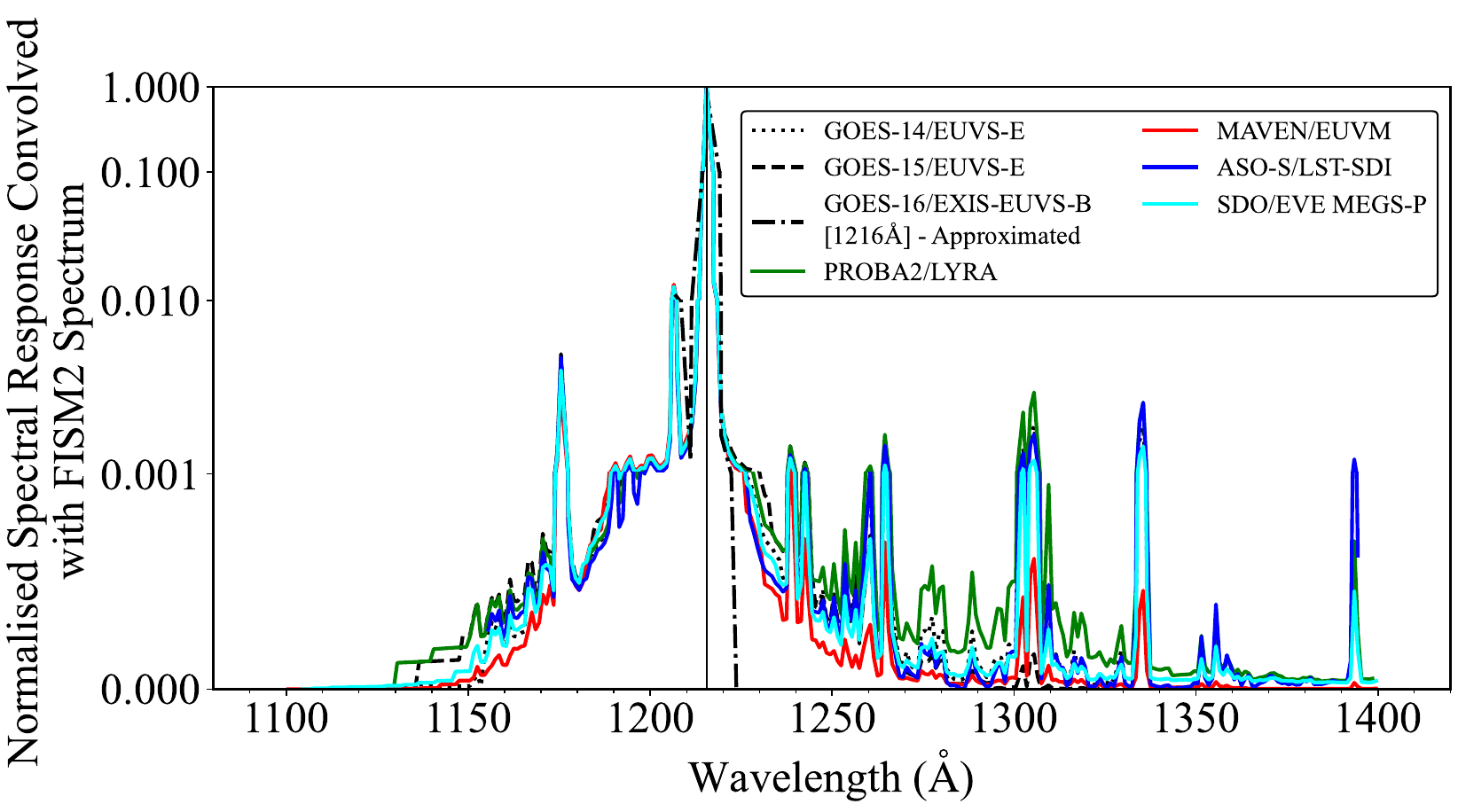}}
\caption{Convolution of a model QS EUV spectrum from FISM2 and the spectral responses for each instrument on a logarithmic scale. The black dot-dashed line denotes a rough approximation of the GOES-16/EXIS-EUVS-B response about the \lya\ cluster derived from the diode responsivities.} 
\label{figure_response_conv}
\end{figure}

The FWHM of the spectral response function for each instrument is presented as a horizontal line in Figure~\ref{figure_bandpass}. Also included is a model of a daily averaged quiet-Sun (QS) EUV spectrum from the \textit{Flare Irradiance Spectral Model v2} (FISM2; \citealt{Chamberlin2020FISM}), where the prominent central peak corresponds to the \lya\ line. The values of the FWHM bandpasses are summarised in Table \ref{table:instrument_summary}. From Figure~\ref{figure_bandpass}, GOES-14/EUVS--E has the broadest bandpass and MAVEN/EUVM has the narrowest. In terms of FWHM values, the values between instruments show reasonable similarity. However, the central point of the bandpass appears to demonstrate a non-negligible difference between each instrument.

To explore the spectral purity of the observations, the spectral response for each instrument was convolved with a model QS spectra from FISM2 in the 1080--1420\AA\ wavelength range (Figure~\ref{figure_response_conv}). From this convolution, it is apparent that for all the instruments examined the measured irradiance across the full bandpass is dominated by \lya. From Figure~\ref{figure_response_conv} it is apparent that contributions to the irradiance from nearby species such as Si~{\sc{iii}} (1206\AA) and O~{\sc{v}} (1218\AA) are found to be 100--1000$\rm \times$ less than that of \lya, thus suggesting it is reasonable to assume that the observed irradiance from each instrument is dominated by \lya\ (\cite{Woods2012EVE} state the filter purity of MEGS-P to be 99\%).

\section{Flare Sample and Analysis} \label{sec:cal&meth}

In order to make instrumental comparison between flare observations, three case studies were conducted each using a unique combination of instruments. The flare sample used for these studies was selected on the fulfillment of the following criteria:

\begin{itemize}

\item Each flare was M-class or above to facilitate sufficient observable \lya\ flux increases attributed to the flare.

\item Each flare was co-observed by the active GOES satellite at the time of the observation and at least one other instrument from Section~\ref{sec:instmeth}.

  \item Each flare occurred on-disk from the respective instrument FOV, reducing the impact of Centre-to-Limb Variation.

  \item The full flare period was observed by each instrument examined in that case study, with a sufficient preflare period to calculate a background flux value.

  \item Instruments must be operating in standard observation modes to prevent additional impact of increased cadences, additional attenuation from filters, spacecraft maneuvers, and variable pointing (imagers, etc.).
  
\end{itemize}

Table~\ref{table:flare_summary} presents a summary of three flares that met the above criteria. The flare timings and classes were taken from the 1--8\AA\ SXR observations from the XRS instrument on the relevant GOES satellite.

\begin{table}[!ht]
\centering
\caption{Observational summary of the flare sample.}
\hspace*{-2.2cm}
\begin{tabular}{lllllll}
\hline
\textbf{Solar} & \textbf{GOES}  & \textbf{GOES}  & \textbf{GOES}   & \textbf{GOES} & \textbf{Heliographic} & \textbf{Observing} \\
\vspace{-0.4cm}
\\
\vspace{0cm}
\textbf{Object} & \textbf{Start}  & \textbf{Peak}  & \textbf{End} & \textbf{Class} & \textbf{Position} & \textbf{Instrument}
\\
\vspace{0cm}
\textbf{Locator} & \textbf{(UT)}  & \textbf{(UT)}  & \textbf{(UT)} & & &
\\
\hline
\vspace{0.2cm}
\FLone & 13:32  & 13:47  & 13:50 & M2.0 & N28W08 & GOES-14/EUVS--E
\\ 
\vspace{0.2cm}
& & & & & & PROBA2/LYRA
\\ 
\vspace{0.2cm}
\FLtwo & 00:14  & 00:29  & 00:39 & M6.7 & N11W60 & GOES-15/EUVS--E
\\ 
\vspace{0.2cm}
& & & & & & MAVEN/EUVM
\\ 
\vspace{0.2cm}
\FLthree & 03:42  & 03:54  & 04:05 & M6.5 & N13W26 & GOES-16/EXIS--EUVS--B
\\ 
\vspace{0.2cm}
& & & & & & ASO-S/LST-SDI
\\ 
\vspace{0.2cm}
& & & & & & SDO/EVE-MEGS-P
\\ 
\end{tabular}
\label{table:flare_summary}
\end{table}

\subsection{Standardisation and Calibration}\label{subsec:standard}

In order to make a reasonable comparison between instruments, \lya\ measurements were converted to disk-integrated irradiance at 1AU, in units of $\rm Wm^{-2}$. The following sections detail the calibration and standardisation procedures conducted for the relevant instruments.

\subsubsection{Scaling Observations to 1AU}\label{subsubsec:Scaling}

As MAVEN is a Mars orbiting mission, a scaling factor is required to convert the \lya\ observations from Mars to Earth distance. This scaling factor is calculated as $\rm(\frac{R_{MS}}{R_{ES}})^{2}$, where $\rm R_{MS}$ and $\rm R_{ES}$ are the Mars-Sun and Earth-Sun separation distances at the time of the flare, respectively. A scaling factor of 2.52 was found for MAVEN. 

Additionally, a light travel-time correction factor was calculated as:

\begin{equation}
\Delta t \ = \ \frac{R_{ScS} - R_{ES}}{c} \rm \ \ seconds    
\label{eq:ltravtime}
\end{equation}

\noindent 
where c is the speed of light in $\rm ms^{-1}$ and the separation distance is given in m. This Earth-Mars travel time could then be subtracted from the observation time to align with observation time taken at 1AU. The calculated value of $\rm \Delta t$ for MAVEN at the time of the flare was found to be 289s. Similar scaling was performed for Earth-orbiting instruments to ensure all observations were scaled to 1AU.

\subsubsection{Radiometric Calibration of Images}\label{radcal}

 Image data require radiometric calibration in order to covert the pixel data in $\rm DNs^{-1}$ to irradiance in $\rm Wm^{-2}$. For ASO-S/LST-SDI, this calibration was carried out using the IDL program \texttt{lst\_radcalib.pro} within the LST analysis package available from the \textit{Science Operation and Data Centre} (SODC). For this calibration, the image data were despiked using ``\textit{L.A.Cosmic}'' cosmic ray identification algorithm (\citealt{vanDokkum2012LAcosmic}), incorporated into the LST software package, and a Radiometric Calibration Factor (RCF) of $\rm 1.98\times10^{-8}\ \rm erg~DN^{-1}~cm^{-2}$ was applied. The RCF is calculated using the ratio of EUVS \lya\ daily averages from GOES to LST data and is time-dependent due to degradation of instrument over the mission lifetime. Following this, image data from ASO-S/LST-SDI was converted from spatially-resolved data to 1-dimensional lightcurves via summation of the pixels present in the full-disk image. It was assumed that the flux within the image exposure time may be extrapolated across the total imaging cadence. This may have an impact on the understanding of the temporal behaviour of the flux, although will not impact the magnitude of \lya\ emission.

 \subsubsection{Centre-to-Limb Variation Correction}\label{CLVcorrection}

\begin{figure}[!ht]    
\centerline{\includegraphics[width=0.45\textwidth]{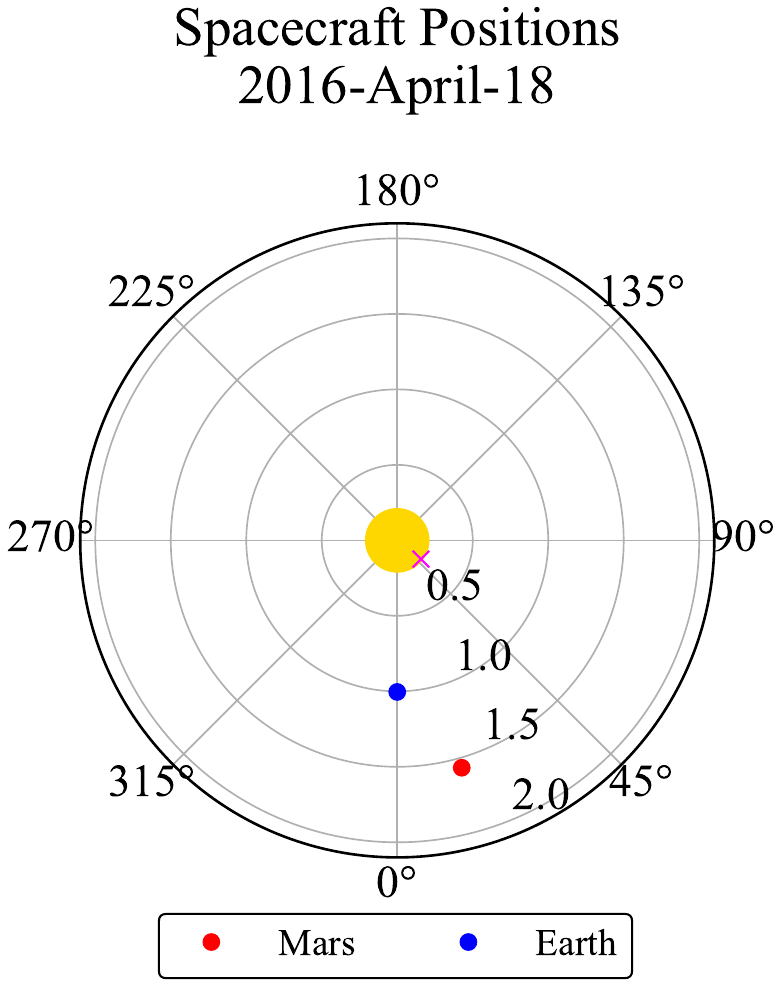}}
\caption{Approximate position relative to the Sun-Earth line for Mars in Heliographic Stonyhurst coordinates on the date of the flare observation. The magenta cross marks the approximate position of \FLtwo. Labelled concentric circles denote distance from the Sun (central yellow circle) in AU.}
\label{figure_spacecraft_pos}
\end{figure}

 Given the position of \FLtwo\ on the solar disk, and the differing vantage points of GOES-15 and MAVEN, a CLV correction factor was required to account for the impact of the relative flare position on the observed \lya\ emission. Firstly, a translation of the longitudinal flare position was performed from the position on-disk as seen from Earth to the approximate position as seen from Mars. This was carried out under the assumption of an unchanged flare latitude via the following: 

 \begin{equation}
\phi_{Mars} \ = \ \phi_{Earth} - \Delta \theta_{Earth-Mars} \ \     
\label{eq:postrans}
\end{equation}

\noindent
where $\phi_{Mars}$ and $\phi_{Earth}$ denote the approximate longitudes as observed from Mars and Earth, respectively, and $\Delta\theta_{Earth-Mars}$ denotes the angular separation between Earth and Mars at the time of the flare. The position of Mars relative to Earth at the time of \FLtwo\ is presented in Figure~\ref{figure_spacecraft_pos}. The value of $\Delta \theta_{Earth-Mars}$ was found to be 15.8$^{\circ}$. From Equation~\ref{eq:postrans}, the angle $\phi_{Mars}$ was calculated as $\sim$44$^{\circ}$, thus giving approximate heliographic coordinates of \FLtwo\ relative to Mars of N11W44.

Following this, a CLV correction factor (C) was calculated for both vantage points using a cosine-squared method, which is justified by the fall-off of irradiance toward the solar limb as demonstrated by \cite{Milligan2020MXFlares}. The CLV correction factor is calculated as:

\begin{equation}
C = \cos^2(\omega) \times \left(1 - \frac{|\phi|}{360}\right) 
\label{eq:CLVcorr}
\end{equation}

\noindent
where $\omega$ is the heliographic latitude in degrees and $\phi$ is the heliographic longitude in degrees. This yielded CLV correction factors for the MAVEN and GOES-15 observations of 0.8030 and 0.8512, respectively. The measured \lya\ fluxes from each instrument were then multiplied by their respective correction factors.

\subsubsection{Recalibration of PROBA2/LYRA Data} \label{subsubsec:recal}

The PROBA2/LYRA observations for \FLone\ were studied in detail by \cite{Kretzshmar2013LyraFLare}. In their study, the authors note that the standard degradation correction may underestimate the flare irradiance increase. Instead they opted to manually calibrate the L1 data, correcting for degradation, dark currents, and spacecraft motion. The \lya\ observations from PROBA2/LYRA were recalibrated using the methods from \cite{Kretzshmar2013LyraFLare} to reproduce their observations for this study. Dark currents were estimated as a function of temperature and removed from the irradiance. A multiplicative degradation correction was applied by dividing the flux by the estimated (approximately linear) degradation for 08--February--2010. Finally, an orbital correction was applied by constructing an orbital pattern for 08--February--2010 and dividing the flux by this. The observations were also degraded to 3s cadence by summation, thus providing an improved signal-to-noise ratio (SNR).

\subsection{Flare Analysis}\label{subsec:flanalysis}

For each flare, the preflare background was taken as the mean flux over a 10 minute period before the XRS start time. The flare contrast was calculated from the peak flux divided by the background, while the excess flux was taken as the background subtracted irradiance. The energy radiated in \lya\ was found by converting the background subtracted irradiance from flux measured at Earth ($\rm I_{Earth}$) to power radiated at the Sun ($\rm P_{Sun}$) by the following:

\begin{equation}
P_{Sun} \ = \ 2\pi R^{2}\times10^{7}~I_{Earth} \ \rm erg~s^{-1}
\label{eqn:wm2toergs}
\end{equation}

\noindent
where $R$ is the Sun-Earth distance and all observations were scaled to 1AU. The value $10^{7}$ is a conversion factor from J to erg. The total energy for each flare was then calculated by integrating the power between the GOES X-ray (XRS 1--8\AA) start and end times. The uncertainty in the observed fluxes was taken as the standard deviation of the flux in the preflare background period. Finally, the time of the measured \lya\ peak for each observation relative to the GOES X-ray peak was calculated as $\rm t^{SXR}_{Peak} - t^{Ly\alpha}_{Peak}$  ($\rm \Delta t_{Peak}$) for each instrument. The uncertainty in peak time was taken as $\rm \pm \frac{cadence}{2}$.

\section{Results}\label{sec:results}

\begin{figure}[!ht]     
\centerline{\includegraphics[width=0.8\textwidth]{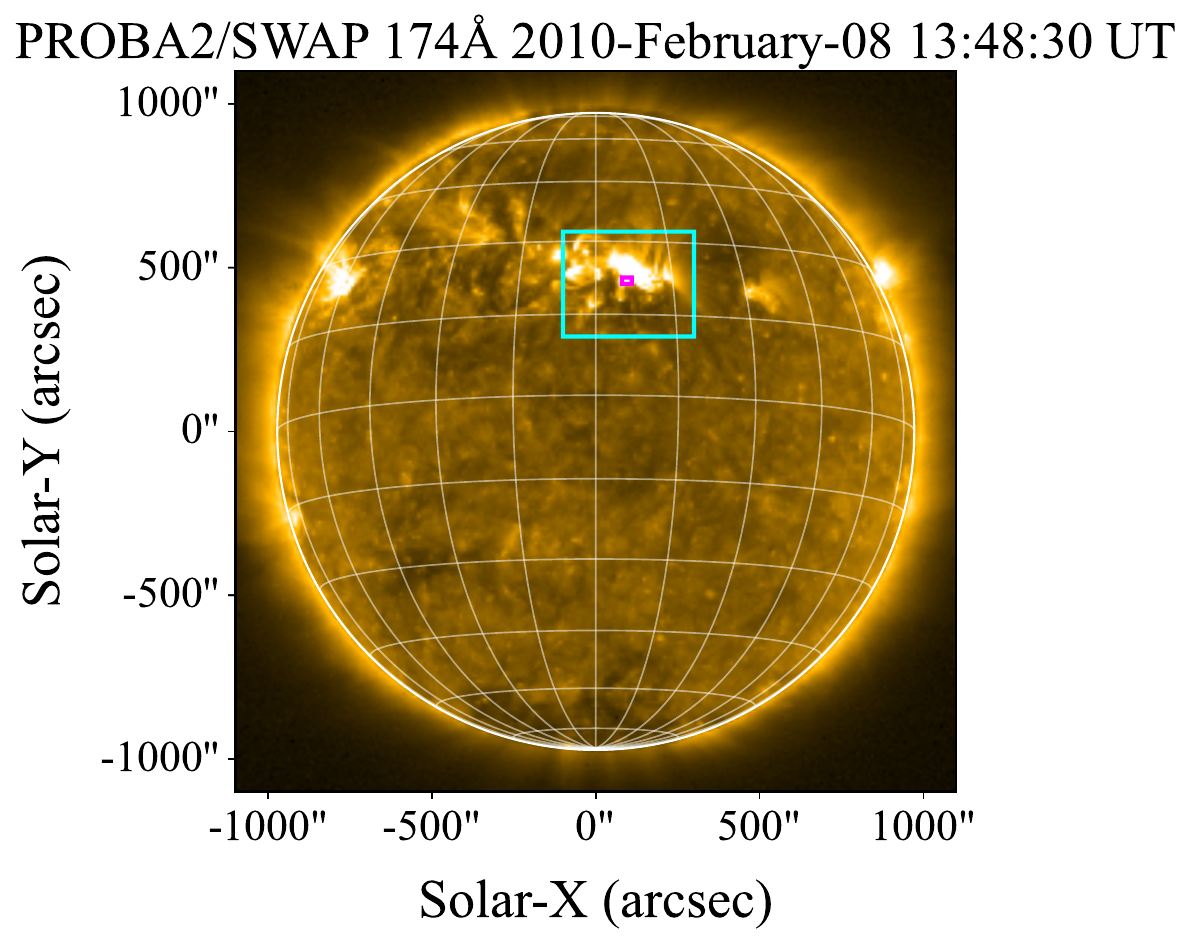}}
\caption{PROBA2/SWAP (174\AA) image of \FLone\ at a time close to the flare peak. The cyan bounding box denotes the active region attributed to the solar flare. The small magenta box denotes the approximate position of the flare source.}
\label{figure_probaimage}
\end{figure}

Each case presented below details flare observations in \lya\ from two or more instruments, examining the relative flux in \lya, the flare-related contrast, the excess flux, energetics, and timing. A quantitative summary of these observations is presented in Table~\ref{table:res_summary}.

\subsection{\FLone}\label{flone}

\begin{figure}[!ht]   
\centerline{\includegraphics[width=1\textwidth]{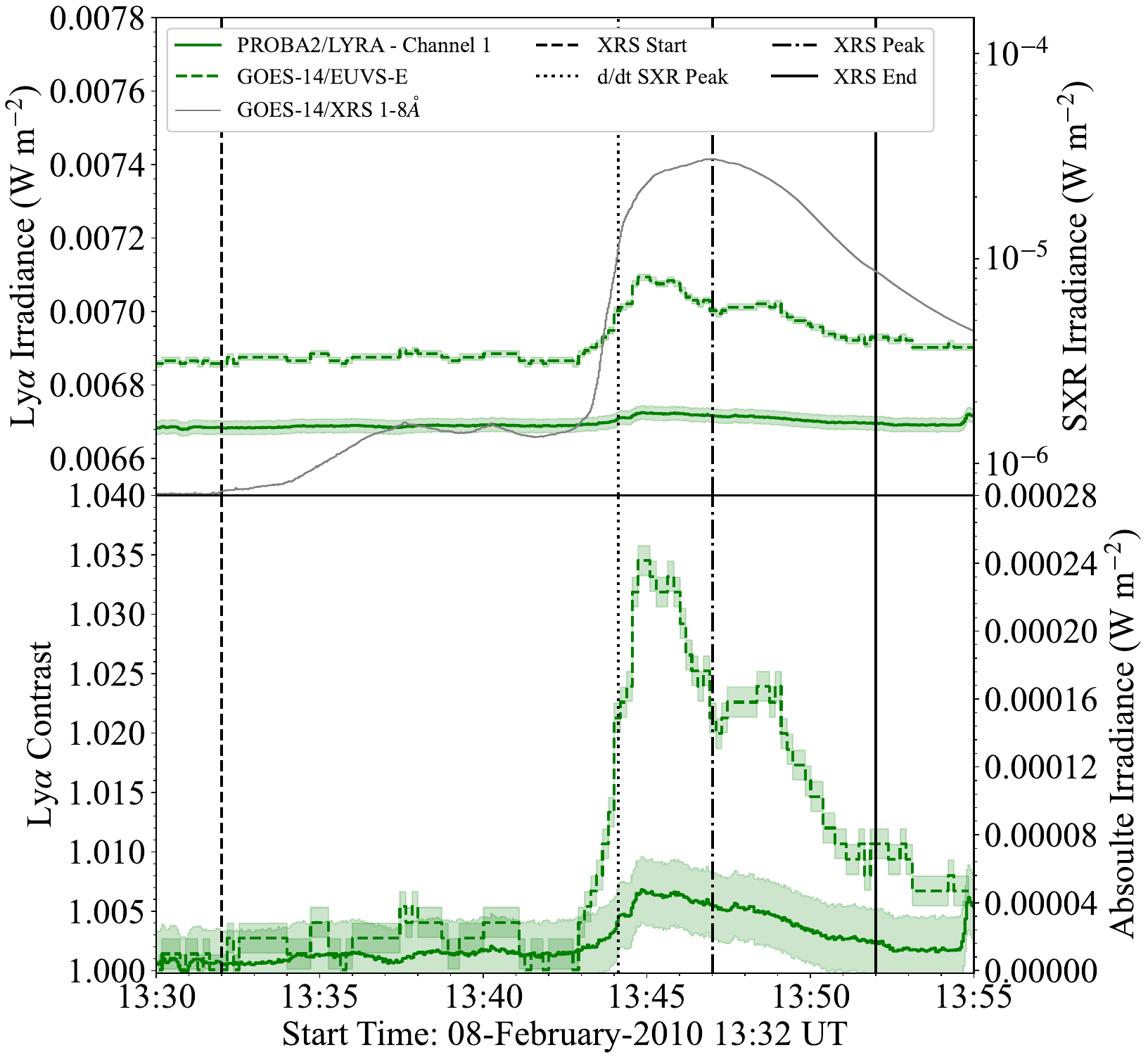}}
\caption{Flare lightcurves from the analysis of \FLone. Top Panel:\lya\ and SXR irradiance lightcurves from GOES-14/EUVS--E (green dashed), GOES-14/XRS (grey solid), and PROBA2/LYRA (green solid). Bottom Panel: Normalised \lya\ enhancement and excess flux of \lya\ emission for each instrument. In both panels, the green dotted line denotes the PROBA2/LYRA data following manual calibration of L1 data. The shaded regions denote the data $\rm \pm \sigma$, where $\rm \sigma$ is the standard deviation of the flux in a 600s preflare window for each observation. Vertical dashed, dot-dashed, and solid lines denote the XRS start, XRS peak, and XRS end times, respectively. Vertical dotted lines denote the peak of the SXR derivative.}
\label{figure_flone_dualPanel}
\end{figure}

The M2.0 flare that occurred on 08--February--2010 was jointly observed by GOES-14/EUVS-E and PROBA2/LYRA. An image from the \textit{Sun Watcher using Active Pixel System detector and Image Processing} (SWAP; \citealt{Berghmans2006SWAP}) onboard PROBA2 is presented in Figure~\ref{figure_probaimage} and the raw lightcurves from both instruments are presented in top panel of Figure~\ref{figure_flone_dualPanel}, while the contrasts and excess fluxes are shown in the bottom panel. From Figure~\ref{figure_flone_dualPanel} it is apparent that there is a substantial disagreement between the relative fluxes from GOES-14/EUVS--E and PROBA2/LYRA, with the flare profile from PROBA2/LYRA appearing notably lower in comparison to GOES-14/EUVS--E. The peak value of the relative flux was found to be $\rm 7.1 \times 10^{-3}~Wm^{-2}$ and $\rm 6.7\times10^{-3}Wm^{-2}$ for GOES/EUVS--E and PROBA2/LYRA, respectively. The respective peaks in \lya\ from the two instruments appear to show temporal agreement with each other but not the SXR derivative. \cite{Kretzshmar2013LyraFLare} conducted a detailed analysis of this flare and found similar results in the flux, suggesting this is due to delayed brightening in the EUV wavelengths as the flare plasma cools and is therefore coronal in origin.

From the bottom panel of Figure~\ref{figure_flone_dualPanel}, the peak contrasts in \lya\ were found to be approximately 3.5\% and 0.7\% for GOES-14/EUVS--E and PROBA2/LYRA, respectively, demonstrating a factor of five difference in their calculated values. Moreover, the peak value of excess flux was found to vary from $\rm 2.4 \times 10^{-4}~Wm^{-2}$ and $\rm 0.5 \times 10^{-4}~Wm^{-2}$ between observations. Converting the excess flux to units of power and integrating over the full flare period, the total energy radiated in \lya\ as observed by GOES-14/EUVS--E was found to be $\rm 1.3 \times 10^{29}~erg$, three times larger than that found for PROBA2/LYRA, which was calculated as $\rm 0.4 \times 10^{29}~erg$. Such a difference in total energy becomes significant to calculations of the chromospheric energy budget when compared with HXR spectroscopic observations. 

One explanation for the significant difference in contrast and flare excess between GOES-14/EUVS-E and PROBA2/LYRA could be contamination of the LYRA bandpass from continuum emission during QS conditions. Specifically the response function of unit 2 of PROBA2/LYRA (used for this study) contains an additional feature at approximately 2000\AA, which is beyond the range of Figure~\ref{figure_bandpass}. The ``out-of-band'' continuum emission in this range accounts for approximately 70\% of the observed QS emission in this unit. Despite this, during flare conditions the enhancements in emission in this spectral range are minimal. Thus, the relative flare signal measured by PROBA2/LYRA in this case can be considered to be dominated by \lya. However, when discussing contrast and flare excess it is likely that the continuum emission in the preflare signal may contribute to a large background flux, which in turn leads to a reduced contrast and flare excess in the overall measurements from PROBA2/LYRA (M. Dominique. 2024 - Private Communication).

\begin{figure}[]    
\centerline{\includegraphics[width=0.7\textwidth]{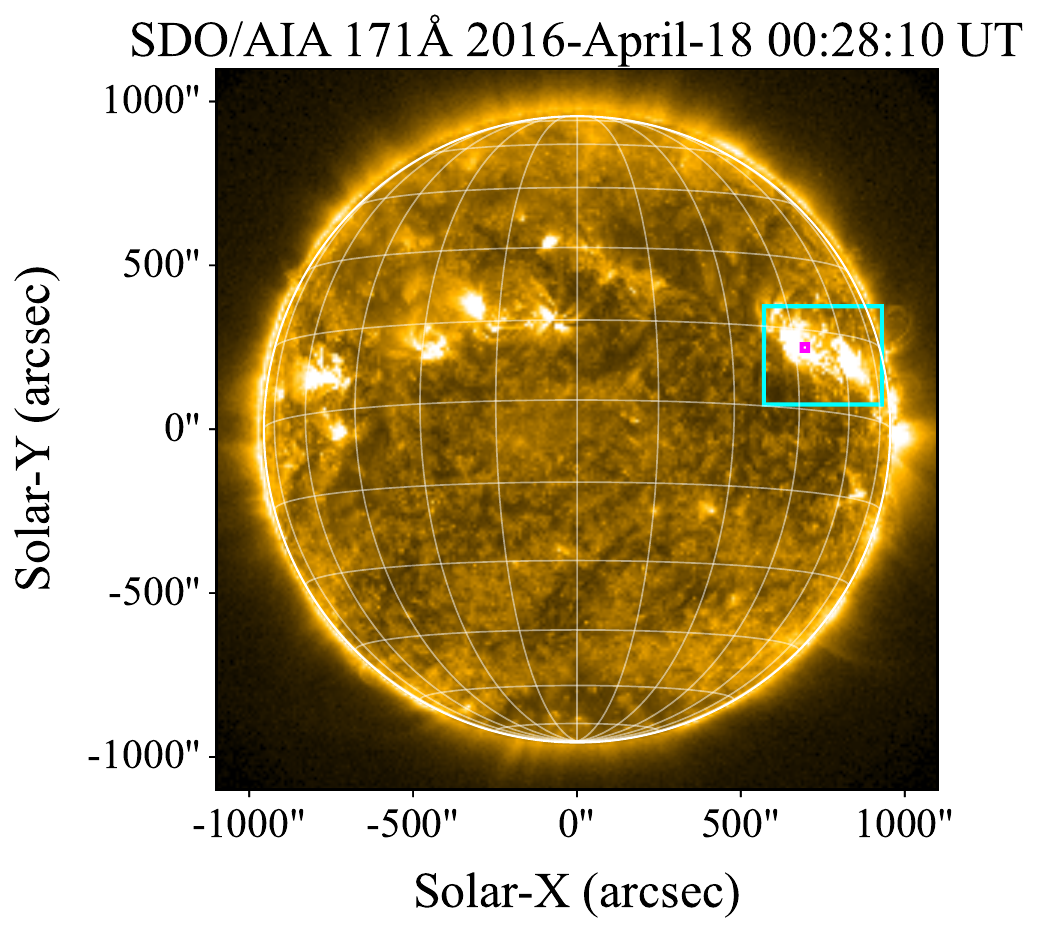}}
\caption{SDO/AIA (171\AA) image of \FLtwo\ at a time close to the flare peak. The cyan bounding box denotes the active region location. The small magenta box denotes the approximate position of the flare source.}
\label{figure_sdo2images}
\end{figure}

 \subsection{\FLtwo}\label{fltwo}

 \begin{figure}[!ht]   
\centerline{\includegraphics[width=0.95\textwidth]{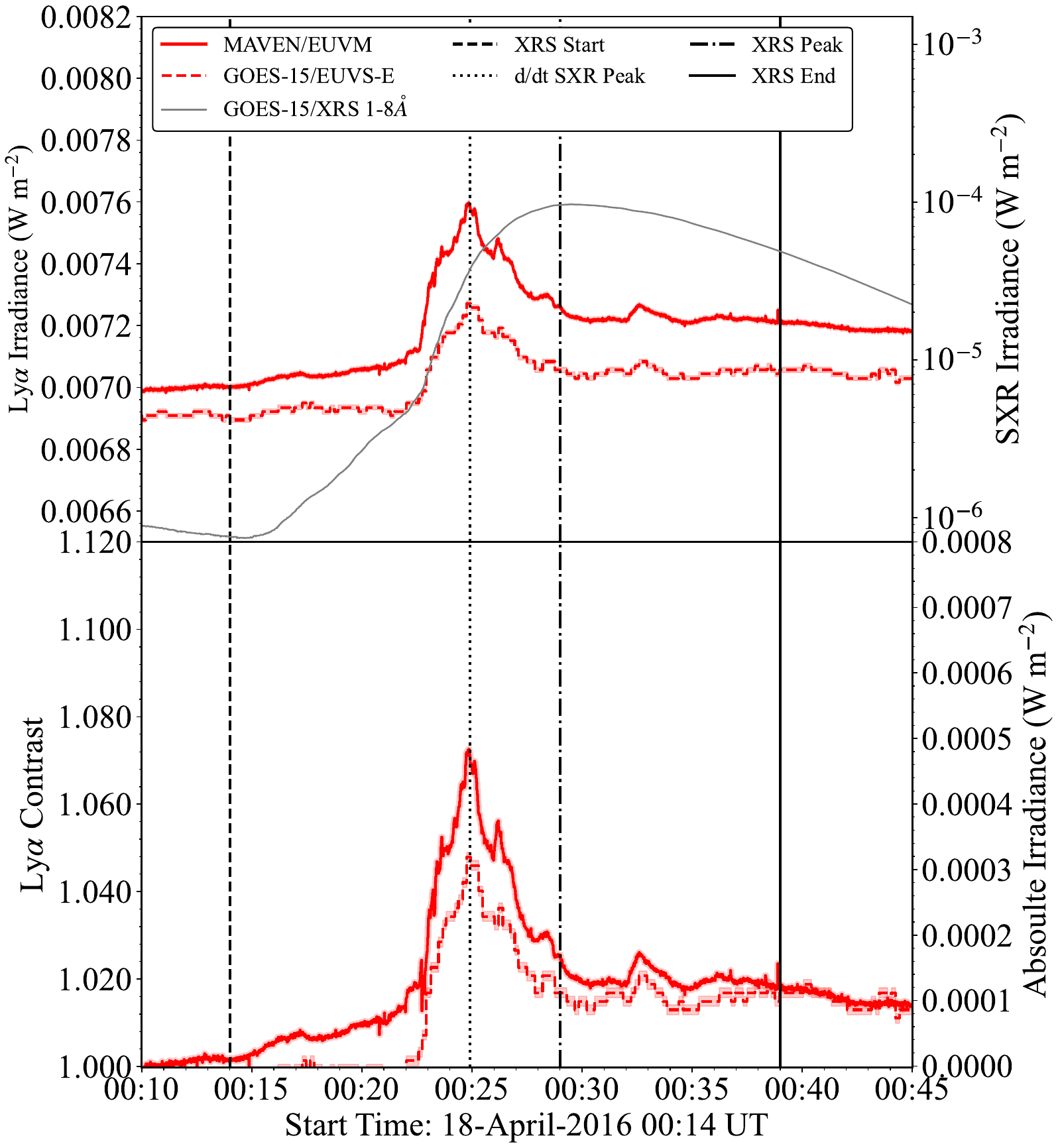}}
\caption{Flare lightcurves from the analysis of \FLtwo. Top Panel: \lya\ and SXR irradiance lightcurves from GOES-15/EUVS--E (red dashed), GOES-15/XRS (grey solid), and MAVEN/EUVM (red solid). Bottom Panel: normalised \lya\ enhancement and excess flux of \lya\ emission for each instrument. The shaded regions denote the data $\rm \pm 1\sigma$, where $\rm \sigma$ is the standard deviation of the flux in a 600s preflare window for each observation. Vertical dashed, dot-dashed, and solid lines denote the XRS start, XRS peak, and XRS end times, respectively. Vertical dotted lines denote the peak of the SXR derivative.}
\label{figure_fltwo_dualPanel}
\end{figure}

The M6.7 flare on 18--April--2016 was co-observed in \lya\ from Earth and Mars by GOES-15/EUVS--E and MAVEN/EUVM, respectively. Imaging for this flare from SDO/AIA (171\AA) is presented in Figure~\ref{figure_sdo2images}. Lightcurves from both observations (following the standardisation of the MAVEN/EUVM data and CLV correction) are presented in the top panel Figure~\ref{figure_fltwo_dualPanel}. The peak relative flux was found to be within 4\% between GOES-15/EUVS--E and MAVEN/EUVM. The observations show remarkable temporal agreement. Slight differences in their temporal behaviour only occur around short bursts, which are captured by MAVEN/EUVM but not GOES-15/EUVS--E due to the factor of 10 difference in their cadences. The contrasts and excess fluxes calculated from each observation are presented in the bottom panel of Figure~\ref{figure_fltwo_dualPanel}. The peak contrasts for GOES-15/EUVS--E and MAVEN/EUVM were found to be 4.2\% and 7.4\%, respectively. The peak values of the excess flux were found to be within a factor of 1.5, with corresponding total energies of $\rm 3.0 \times 10^{29}~erg$ and $\rm 5.5 \times 10^{29}~erg$. Thus, the energy found using MAVEN/EUVM is almost a factor of two larger than that of GOES-15/EUVS--E.

\begin{figure}[!ht]     
\centerline{\includegraphics[width=0.7\textwidth]{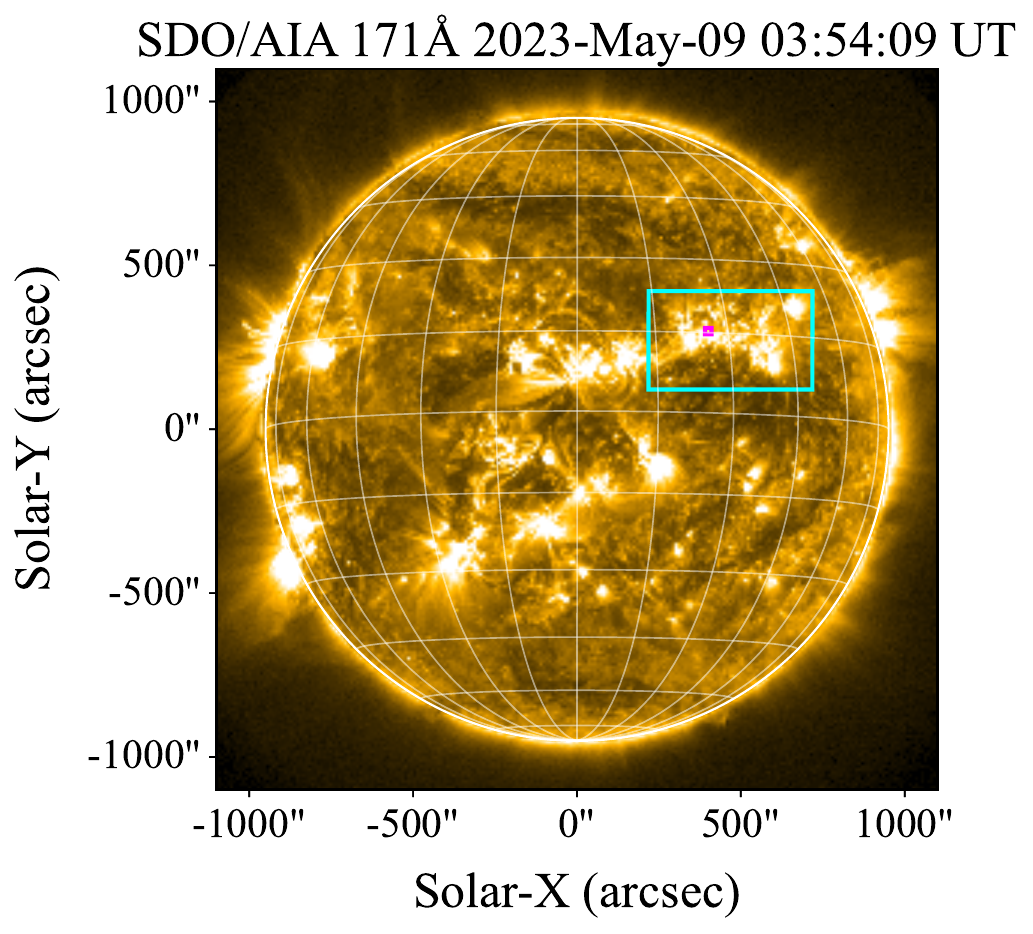}}
\caption{SDO/AIA (171\AA) image of \FLthree\ at a time close to the flare peak. The cyan bounding box denotes the approximate flare location. The small magenta box denotes the approximate position of the flare source.}
\label{figure_sdo4image}
\end{figure}

\begin{figure}[!ht]   
\centerline{\includegraphics[width=0.95\textwidth]{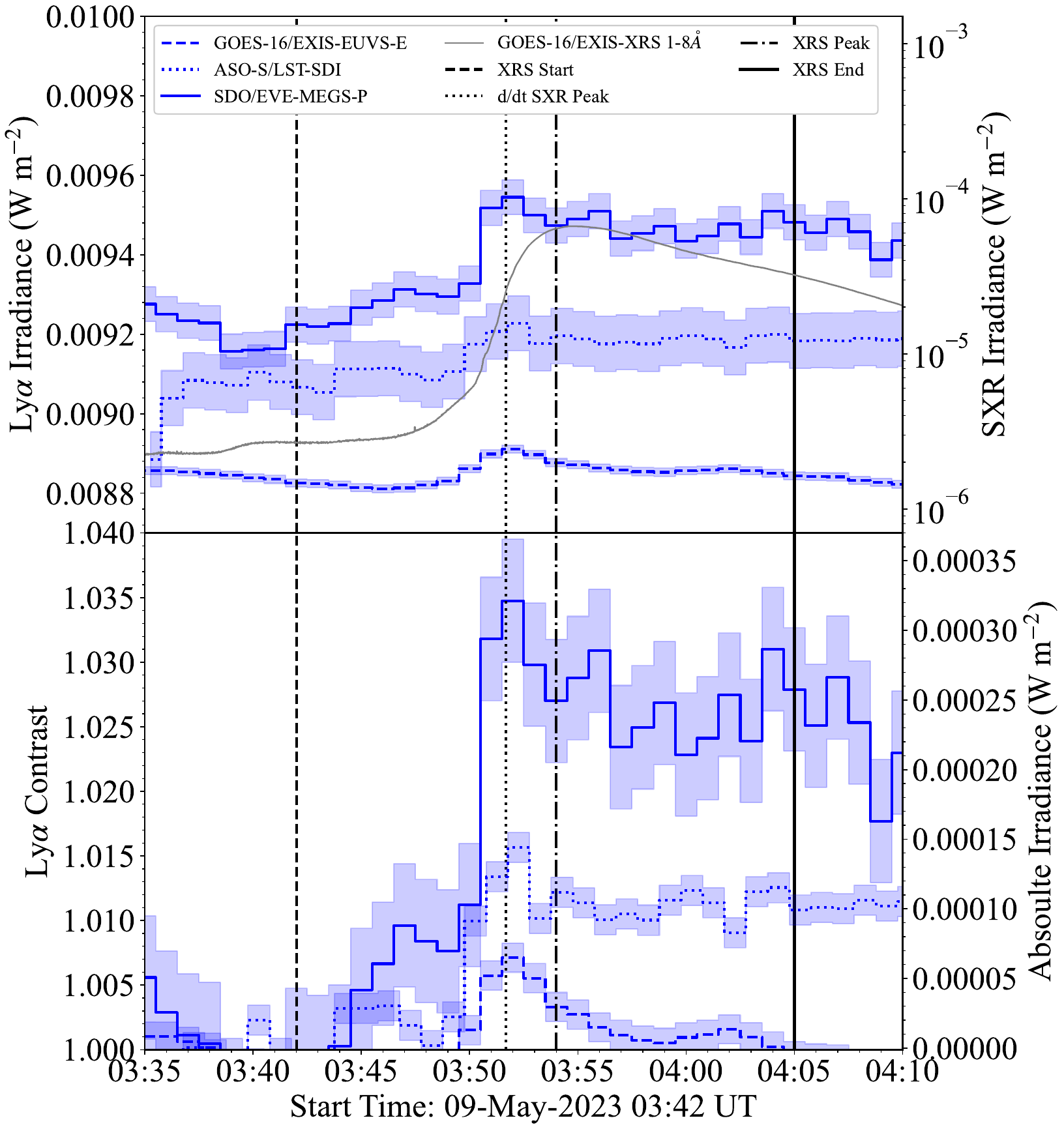}}
\caption{Flare lightcurves from the analysis of \FLthree. Top Panel:\lya\ and SXR irradiance lightcurves from GOES-16/EXIS--EUVS--B (blue dashed), GOES-16/EXIS--XRS (grey solid), ASO-S/LST-SDI (blue dotted), and SDO/EVE-MEGS-P (blue solid). Bottom Panel: normalised \lya\ enhancement and excess flux of \lya\ emission for each instrument. The shaded regions denote the data $\rm \pm \sigma$, where $\rm \sigma$ is the standard deviation of the flux in a 600s preflare window for each observation. Vertical dashed, dot-dashed, and solid lines denote the XRS start, XRS peak, and XRS end times, respectively. Vertical dotted lines denote the peak of the SXR derivative.}
\label{figure_FLthree_dualPanel}
\end{figure}

\subsection{\FLthree}

The M6.5 flare on 09--May--2023 was observed photometrically in \lya\ by both GOES-16/EXIS--EUVS--B and SDO/EVE-MEGS-P, as well as imaged by ASO-S/LST-SDI. A context image from SDO/AIA (171\AA) is presented in Figure~\ref{figure_sdo4image}. Following the radiometric calibration of the ASO-S/LST-SDI data detailed in Section~\ref{radcal}, comparison was made with the photometric observations. The top panel of Figure~\ref{figure_FLthree_dualPanel} presents the relative flux from each instrument. It is apparent that peak relative fluxes for the three observing instruments differ by approximately 6\%. The larger flux found SDO/EVE-MEGS-P may be attributed to broadening of the instrument bandpass over time, which is not necessarily measurable (\textit{Woodraska. 2023 - Private Communication}).

Despite the similarity in flare profiles, the contrasts were found to range from 0.7\% to 3.5\%. This suggests that despite the similarity in the measured \lya\ peak, the sensitivity to the QS flux levels may drive discrepancies in the background values with which the contrasts are calculated. Similarly, the peak excess fluxes differed by up to $\rm 2.5\times 10^{-4}~Wm^{-2}$ between all observations, translating to a maximum discrepancy in total energy of $\rm 4.0 \times 10^{29}~erg$. 

\begin{figure}[!ht]  
\centerline{\includegraphics[width=0.9\textwidth]{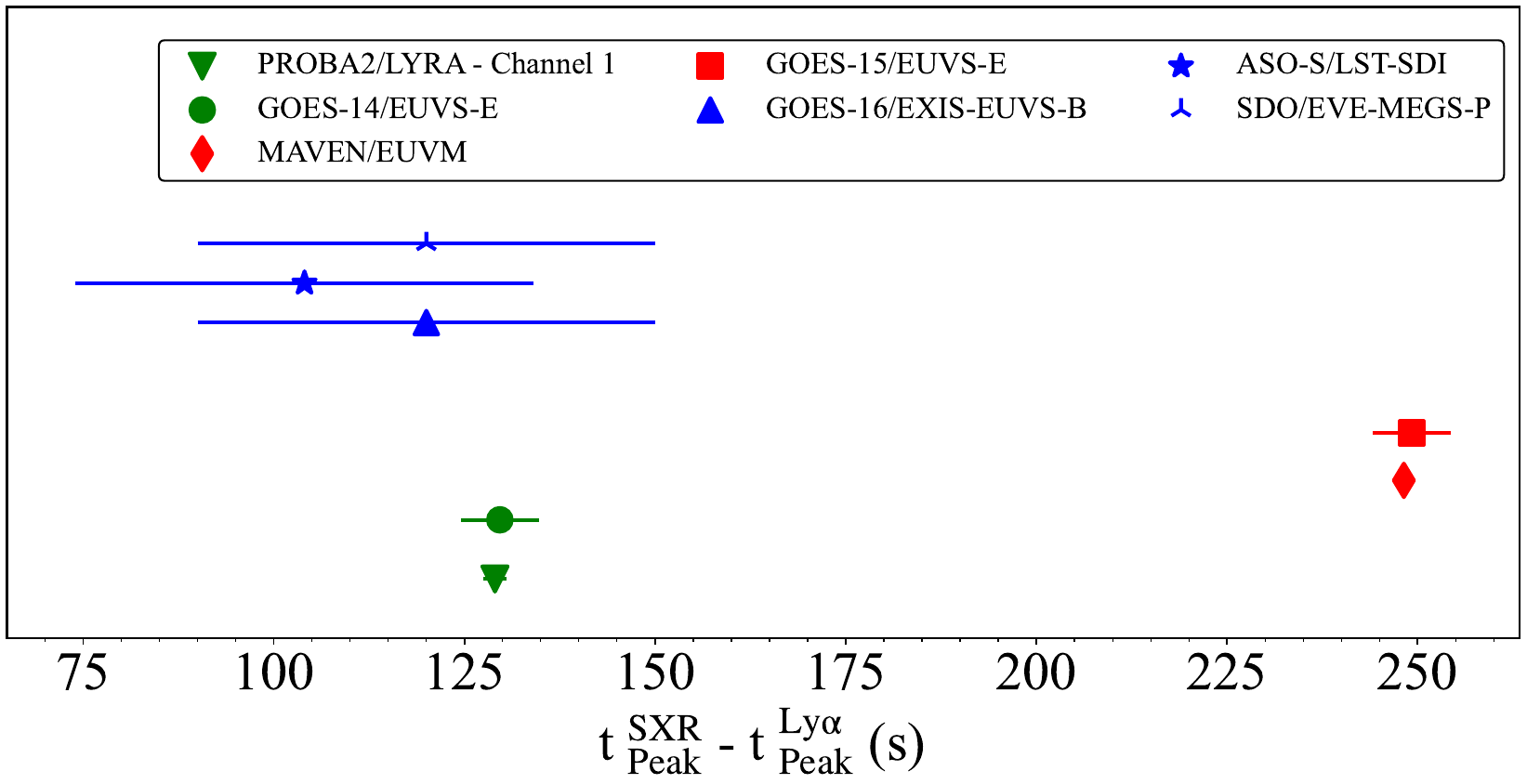}}
\caption{Time differences between the SXR peak (1--8\AA) from GOES and the \lya\ peak for each observing instrument for each flare. Colours denote each flare in the sample. Markers denote the unique observing instrument. The width of the horizontal lines affixed to each marker correspond to the instrument cadence.}
\label{figure_peaktime}
\end{figure}

\subsection{Timings} \label{subsec:timing}

For each flare, the time difference between the GOES SXR peak in the 1--8\AA\ band and the observed \lya\ peak from each instrument was calculated in order to identify any lag or lead times. Figure~\ref{figure_peaktime} presents the absolute value of $\rm t^{SXR}_{Peak} - t^{Ly\alpha}_{Peak}$  ($\rm \Delta t_{Peak}$) for each instrument. The uncertainty is taken as $\rm \pm \frac{cadence}{2}$, given by the horizontal error bars. For \FLone\ there is a disparity of 0.6s between the $\rm \Delta t_{Peak}$ values calculated for PROBA2/LYRA and GOES-14/EUVS--E. There is a remarkable agreement between the $\rm \Delta t_{Peak}$ values calculated for GOES-15/EUVS--E and MAVEN/EUVM for \FLtwo\ with a discrepancy of 0.3s. For \FLthree, there is a broad agreement in $\rm \Delta t_{Peak}$ between GOES-16/EXIS--EUVS--B and SDO/EVE-MEGS-P, although the uncertainty of these values is particularly large due to the relatively low (60s) cadence between measurements. There is a time difference of 16s in $\rm \Delta t_{Peak}$ between these two instruments and ASO-S/LST-SDI. However, the $\rm \Delta t_{Peak}$ values for this flare all lie withing the associated uncertainties of one another. Generally, the peak times between instruments are in relatively good agreement with each other given the associated uncertainties in the recorded times.

\begin{sidewaystable}[ht]
\centering
\caption{Summary of the analysis results for each observation. Relative flux peak values were taken from calibrated irradiances that were scaled to 1AU and given to the nearest 2 significant figures.  \label{table:res_summary}}
\begin{tabular}{lllllll}
\hline
\textbf{Flare} &  \textbf{Observing} & \textbf{Peak Flux$\rm_{Rel}$} & \textbf{Peak} & \textbf{Peak Flux$\rm_{Exc}$} & \textbf{Total Energy}  & \textbf{$\rm t^{SXR}_{Peak} - t^{Ly\alpha}_{Peak}$} \\
\vspace{-0.4cm}
\\
\vspace{0cm}
\textbf{Identifier} & \textbf{Instrument} & \textbf{($\rm 10^{-3} Wm^{-2}$)} & \textbf{Contrast (\%)} & \textbf{($\rm 10^{-4} Wm^{-2}$)} & \textbf{($\rm 10^{29} erg$)} & \textbf{(s)}
\\
\hline
\vspace{0.2cm}
\FLone  & GOES-14/EUVS--E  & 7.1 $\pm$ 0.009 & 3.5 $\pm$ 0.1 & 2.4 $\pm$ 0.09 & 1.3 $\pm$ 0.001 & 129.6$\pm$5.12
\\ 
\vspace{0.2cm}
 & PROBA2/LYRA  & 6.7 $\pm$ 0.01 & 0.7 $\pm$ 0.2 & 0.5 $\pm$ 0.1 & 0.4 $\pm$ 0.00007 & 129.0$\pm$1.5
\\ 
\vspace{0.2cm}
\FLtwo  & GOES-15/EUVS--E & 7.3 $\pm$ 0.008 & 4.4 $\pm$ 0.1 & 3.2 $\pm$ 0.08 & 3.0 $\pm$ 0.001 & 249.2$\pm$5.12
\\ 
\vspace{0.2cm}
 & MAVEN/EUVM & 7.6 $\pm$ 0.007 & 7.4 $\pm$ 0.1 & 5.0 $\pm$ 0.007 & 5.5 $\pm$ 0.0001 & 248.1$\pm$0.5
\\ 
\vspace{0.2cm}
\FLthree  & GOES-16/EXIS--EUVS--B & 8.9 $\pm$ 0.01 & 0.7 $\pm$ 0.1 & 0.7 $\pm$ 0.01 & 0.3 $\pm$ 0.009 & 120.0$\pm$30.0
\\ 
\vspace{0.2cm}
&  SDO/EVE-MEGS-P & 9.5 $\pm$ 0.04 & 3.5 $\pm$ 0.5 & 3.2 $\pm$ 0.04 & 4.3 $\pm$ 0.04 & 120.0$\pm$30.0
\\ 
\vspace{0.2cm}
& ASO-S/LST-SDI & 9.2 $\pm$ 0.01 & 1.6 $\pm$ 0.1 & 1.4 $\pm$ 0.01 & 2.0 $\pm$ 0.009 & 104.0$\pm$30.0
\\ 
\end{tabular}
\end{sidewaystable}

\section{Discussion and Future Missions} \label{sec:Discussion}

In this work, we present an inter-instrument comparison of flare-related \lya\ observations for three M-class flares, focussing on the relative and excess flux, the calculated flare contrast, and the total radiated energy. From this we find measurement inconsistencies to exist to a varying degree across seven different \lya\ instruments. The key findings are as follows:

\begin{itemize}

\item Relative fluxes between instruments for all three flares examined here are in sufficient agreement, such that the conclusions drawn from their analysis will be negligibly impacted by the choice of instrument.

\item Discrepancies in the contrasts and excess fluxes are of sufficient degree to have a substantial effect on the conclusions drawn from multi-instrument \lya\ studies. 

\item The calculated energies between instruments are sufficiently different to potentially have a significant impact on constraining the contribution of \lya\ to the radiated energy budget of solar flares. 

\item The flare timings between instruments are in relatively good agreement considering the uncertainties and therefore are minimally impacted by the choice of observing instrument. 
  
\end{itemize}

\cite{Milligan2020MXFlares} demonstrated that 95~\% of M- and X-class flares observed by GOES-15/EUVS--E have an associated \lya\ contrast of $\leq$10\%, with an upper limit to the contrast of $\sim$30~\%, while \cite{Raulin2013LyaIonosphere} found \lya\ contrasts measured by PROBA2/LYRA to be consistently $<$1\%. The flares examined here were found to have contrasts ranging from 0.3\% up to 7.4\%, but importantly it has been demonstrated here that these ranges may be influenced by the observing instrument. The findings presented here show that the contrast range may vary by up to $\sim$3.5\% depending on the observing instrument. This discrepancy is over half of the range of contrast variability found in \cite{Greatorex2023EquivMagFlares}. Moreover, it has been shown that the calculated values of excess flux can vary by up to a factor of five between instruments. This implies that analyses of solar flare excess in \lya\ may be significantly impacted by the choice of observing instrument, thus influencing the conclusions drawn from flare observations and further calculations of the flare energy carried out using these fluxes. On the contrary, the values of relative flux were found to be in sufficient agreement such that any analysis done with these fluxes would be minimally impacted by the discrepancies between observations from different instruments. 

Calculations of the radiated energy in \lya\ are also vital for estimating the contribution of this wavelength (and therefore others) to the radiated energy budget of the chromosphere. Relatively few studies have compared the radiated energy in \lya\ to the incident nonthermal electron energy deposited into the chromosphere during flares. From those studies, \lya\ has been found to radiate up to $\sim$8\% of the chromospheric energy (\citealt{Milligan2014EUVEnergy}). Here it has been demonstrated that calculations of the radiated energy in \lya\ may vary by up to an order of magnitude between observations from different instruments. This has significant implication for statistical studies attempting to quantify the radiated energy in \lya\ over a large flare sample, particularly when using multiple instruments as part of those observations. In this work, it has not been possible to compute the nonthermal electron energies for each flare due to a lack of uniform HXR observations from a single observing instrument. Introducing variation in HXR observations will add unnecessary excess uncertainty in calculated energies that may distract from the discrepancies in the \lya\ observations. Under the assumption of a single nonthermal energy for each flare, the discrepancies in the calculated \lya\ energy could substantially alter estimates of the percentage contribution of \lya\ emission to the radiated energy budget of the chromosphere.

 Data driven models such as FISM2 depend on observations to form statistics, from which approximations can be made for emission spectra from solar flares. These models have application in the study of atmospheric responses to solar flares (\citealt{Qian2010FISMApplication, Qian2011FISMApplication, Qian2012FISMApplication, Lollo2012FISMApplicationMars}), as well as in the development of models designed to aid in the analysis of observations from missions such as MAVEN (\citealt{Thiemann2017MavenEUVMModels, Chaffin2015MAVEN-FISM-MApplication, Chaufray2015MAVEN-FISM-MApplication, Jain2015MAVEN-FISM-MApplication, Sakai2016MAVEN-FISM-MApplication}). The interpretation of the results from studies employing these models is thus reliant on the precision and credibility of the underlying observations. 

Significant disagreement in the observed \lya\ peak times between GOES/EUVS--E and SDO/EVE-MEGS-P were presented by \cite{Milligan2016SDO_EVE_MEGSP}. They found a delay in peak times between instruments of 5--10 minutes, which was eventually attributed to the Kalman filter used to smooth data in SDO/EVE processing; this filter was subsequently replaced with a Fourier transform filter, which ultimately prevented the recurrence of this issue. The time discrepancies presented in this work are not as substantial as those presented in \cite{Milligan2016SDO_EVE_MEGSP}; the differences in peak times between instruments could be attributed to the variability in cadences, or in the case of imagers, the assumptions of constant flux between exposure, which may be an oversimplification of the true behaviour of the \lya\ flux. While temporal discrepancies between observations may not have a substantial impact on calculations flare energetics, they may affect the understanding of energy transport processes in flares as well as the derived relationships between solar flares and associated atmospheric responses in the unique sub-regions of the ionosphere (\citealt{Berdermann2018SpaceWeatherNavigation, Raulin2013LyaIonosphere, Hayes2021Ionosphere, Milligan2020MXFlares, Chakraborty2021IonosphericSluggishness, Barta2022IonosphericResponseFlares}). 

The convolution of the FISM2 spectra in Section~\ref{sec:bpsrf} demonstrates that there is minimal contribution to the total measured \lya\ irradiance from additional species in the bandpass of each instrument. Minor variations between instruments are present in the wings of the \lya\ profile as well as in the blueward and redward wavelengths surrounding the \lya\ line. However, the core emission significantly dominates the measured flux in all cases by at least two orders of magnitude, therefore the total measured irradiance may be considered to be predominantly from the \lya\ line. 

Given the standardisation of the observations carried out in this work, it is implicit that the observed discrepancies found between instruments are unlikely to be attributable to the observation field of view or observing techniques of the instruments. Instead, the discrepancies in the measured flux between instruments may be driven by inherent properties of the instrument itself or the calibration process carried out on the raw data. Absorption from the goecorona may be able to account for differences in observed flux between instruments at different orbital heights. Specifically, satellites in Low-Earth Orbit may measure lower levels of \lya\ irradiance compared to those with more extended orbits due to the absorption of \lya\ photons by Hydrogen in the Earth's geocorona. \cite{Wauters2022M67Lya} suggested that geocornal absorption may be a potential explanation for a factor of 20 difference between GOES-15/EUVS-E and PROBA2/LYRA observations of a solar flare in \lya. Additionally, the spectral assumptions used to scale observations for each instrument may influence the returned flare profiles. For example, GOES-15/EUVS-E data are scaled to the \textit{Whole Heliosphere Interval} QS reference spectrum and thus systematic uncertainties may be present in the observed flare data (\citealt{Woods2009WHI, Milligan2021B&CClassFlares}).

Ultimately, it is important to acknowledge any discrepancy in observations between instruments, particularly when conducting multi-instrument studies, which are becoming evermore possible with the expanding availability of flare-related \lya\ observations. Previously, GOES observations in the SXRs have been considered the ``industry standard'' for flare classifications and timings. It may be of value to establish some form of agreeable standard of observation to scale measurements from dedicated \lya\ missions to. Analysis of results from upcoming missions with \lya\ observing capabilities such as Solar-C featuring the Solar Spectral Irradiance Monitor (SoSpIM; \citealt{Harra2022SoSPIMPreceedings}), and the \textit{The Solar eruptioN Integral Field Spectrograph} (SNIFS; \citealt{Chamberlin2020SNIFS}) sounding rocket will benefit from recognition of the diversity of conclusions that may be drawn from \lya\ observations depending on the observing instrument. The findings presented here may guide interpretation of the observations taken by the current and new generation of \lya\ instruments as part of future studies. 

\section{Acknowledgements}
H.J.G would like to thank the UK’s Science \& Technology Facilities Council (ST/W507751/1) for supporting this research. R.O.M would like to thank the UK’s Science \& Technologies Facilities Council for the award of an Ernest Rutherford Fellowship (ST/N004981/2). They also thank M. Mathioudakis (QUB) for stimulating discussions on this work. H.J.G would also like to thank the attendees of the 2024 SoSpIM meeting for their feedback on the results presented in this work. The authors would like to thank Y. Li (PMO) and the LST Team for their invaluable assistance with the ASO-S data used for this study.  The authors would also like to thank J. Machol (CU-CIRES \& NOAA-NCEI) for their useful comments on the discussion of the GOES mission. They also thank M. Dominique (ROB) for their invaluable assistance with the PROBA2 data used in this study. Furthermore, they thank P. Chamberlin, T. Woods, and D. Woodraska (CU-LASP) for their help and advice with both the MAVEN and SDO data incorporated into this work. The ASO-S mission is supported by the Strategic Priority Research Program on Space Science, the Chinese Academy of Sciences (CAS). LYRA is a project of the Centre Spatial de Liege, the Physikalisch-Meteorologisches Observatorium Davos and the Royal Observatory of Belgium funded by the Belgian Federal Science Policy Office (BELSPO) and by the Swiss Bundesamt für Bildung und Wissenschaft. Finally, the authors would like to thank the referee of this article for their useful comments, which have greatly contributed to the overall improvement of this work. 

\section{Data Availability}

No new data was generated as part of this study. Observation and responsivity data are available for GOES-16 from \url{https://www.ncei.noaa.gov/products/goes-r-extreme-ultraviolet-xray-irradiance} and for GOES-14 and-15 from \url{https://www.ncei.noaa.gov/products/goes-1-15/space-weather-instruments}.
PROBA2 data is available from the LYRA data centre, or directly here: \url{https://proba2.sidc.be/data/LYRA}. The response function data for LYRA was privately communicated to the authors by M. Dominique and is presented in \cite{Dominique2013LYRAInstr}. MAVEN/EUVM data can be accessed through the Planetary Data System, or directly here: \url{https://pds-ppi.igpp.ucla.edu/mission/MAVEN/Extreme_Ultraviolet_Monitor}. SDO/EVE data may be accessed via the CU/LASP EVE data site or directly here: \url{https://lasp.colorado.edu/eve/data_access/index.html}, the response function for MEGS-P was privately communicated by T. Woods. ASO-S/LST-SDI data is available from the SODC via \url{http://aso-s.pmo.ac.cn/sodc/dataArchive.jsp}. ASO-S analysis software may be accessed via \url{http://aso-s.pmo.ac.cn/sodc/analysisSoftware.jsp}. The response function of ASO-S/LST-SDI was privately communicated to the authors by Y. Li and is presented in \cite{Chen2024LST}. The FISM2 data are publicly available on the LISIRD data site on the FISM page, or directly here: \url{http://lasp.colorado.edu/lisird/}. All observation data in this study may be analysed using SolarSoftware (SSWIDL; \url{https://www.lmsal.com/solarsoft/}) or SunPy (\url{https://sunpy.org/}; \citealt{SunPy2020}).

%

%



%
%

%
%
%
%
%
%
%

%
%
\bibliographystyle{plainnat}  
\bibliography{references}  

\begin{thebibliography}{67}
\providecommand{\natexlab}[1]{#1}
\providecommand{\url}[1]{\texttt{#1}}
\expandafter\ifx\csname urlstyle\endcsname\relax
  \providecommand{\doi}[1]{doi: #1}\else
  \providecommand{\doi}{doi: \begingroup \urlstyle{rm}\Url}\fi

\bibitem[Barta et~al.(2022)Barta, Natras, Srećković, Koronczay, Schmidt, and Šulic]{Barta2022IonosphericResponseFlares}
Veronika Barta, Randa Natras, Vladimir Srećković, David Koronczay, Michael Schmidt, and Desanka Šulic.
\newblock Multi-instrumental investigation of the solar flares impact on the ionosphere on 05–06 december 2006.
\newblock \emph{Frontiers in Environmental Science}, 10, 2022.
\newblock ISSN 2296-665X.
\newblock \doi{10.3389/fenvs.2022.904335}.
\newblock URL \url{https://www.frontiersin.org/articles/10.3389/fenvs.2022.904335}.

\bibitem[{Bartoe} et~al.(1977){Bartoe}, {Brueckner}, {Purcell}, and {Tousey}]{Bartoe1977NRLSpec}
J.~D.~F. {Bartoe}, G.~E. {Brueckner}, J.~D. {Purcell}, and R.~{Tousey}.
\newblock {Extreme ultraviolet spectrograph ATM experiment S082B.}
\newblock \emph{American Journal}, 16:\penalty0 879--886, April 1977.
\newblock \doi{10.1364/AO.16.000879}.

\bibitem[{BenMoussa} et~al.(2015){BenMoussa}, {Giordanengo}, {Gissot}, {Dammasch}, {Dominique}, {Hochedez}, {Soltani}, {Bourzgui}, {Saito}, {Sch{\"u}hle}, {Gottwald}, {Kroth}, and {Jones}]{BenMoussa2015LyraDeg}
A.~{BenMoussa}, B.~{Giordanengo}, S.~{Gissot}, I.~E. {Dammasch}, M.~{Dominique}, J.~F. {Hochedez}, A.~{Soltani}, N.~{Bourzgui}, T.~{Saito}, U.~{Sch{\"u}hle}, A.~{Gottwald}, U.~{Kroth}, and A.~R. {Jones}.
\newblock {Degradation assessment of LYRA after 5 years on orbit - Technology Demonstration -}.
\newblock \emph{Experimental Astronomy}, 39\penalty0 (1):\penalty0 29--43, March 2015.
\newblock \doi{10.1007/s10686-014-9437-7}.

\bibitem[Berdermann et~al.(2018)Berdermann, Kriegel, Banyś, Heymann, Hoque, Wilken, Borries, Heßelbarth, and Jakowski]{Berdermann2018SpaceWeatherNavigation}
J.~Berdermann, M.~Kriegel, D.~Banyś, F.~Heymann, M.~M. Hoque, V.~Wilken, C.~Borries, A.~Heßelbarth, and N.~Jakowski.
\newblock Ionospheric response to the x9.3 flare on 6 september 2017 and its implication for navigation services over europe.
\newblock \emph{Space Weather}, 16\penalty0 (10):\penalty0 1604--1615, 2018.
\newblock \doi{https://doi.org/10.1029/2018SW001933}.
\newblock URL \url{https://agupubs.onlinelibrary.wiley.com/doi/abs/10.1029/2018SW001933}.

\bibitem[{Berghmans} et~al.(2006){Berghmans}, {Hochedez}, {Defise}, {Lecat}, {Nicula}, {Slemzin}, {Lawrence}, {Katsyiannis}, {van der Linden}, {Zhukov}, {Clette}, {Rochus}, {Mazy}, {Thibert}, {Nicolosi}, {Pelizzo}, and {Sch{\"u}hle}]{Berghmans2006SWAP}
D.~{Berghmans}, J.~F. {Hochedez}, J.~M. {Defise}, J.~H. {Lecat}, B.~{Nicula}, V.~{Slemzin}, G.~{Lawrence}, A.~C. {Katsyiannis}, R.~{van der Linden}, A.~{Zhukov}, F.~{Clette}, P.~{Rochus}, E.~{Mazy}, T.~{Thibert}, P.~{Nicolosi}, M.~G. {Pelizzo}, and U.~{Sch{\"u}hle}.
\newblock {SWAP onboard PROBA 2, a new EUV imager for solar monitoring}.
\newblock \emph{Advances in Space Research}, 38\penalty0 (8):\penalty0 1807--1811, January 2006.
\newblock \doi{10.1016/j.asr.2005.03.070}.

\bibitem[Boerner et~al.(2012)Boerner, Edwards, Lemen, Rausch, Schrijver, Shine, Shing, Stern, Tarbell, Title, Wolfson, Soufli, Spiller, Gullikson, McKenzie, Windt, Golub, Podgorski, Testa, and Weber]{Boerner2012AIA}
Paul Boerner, Christopher Edwards, James Lemen, Adam Rausch, Carolus Schrijver, Richard Shine, Lawrence Shing, Robert Stern, Theodore Tarbell, Alan Title, C~Jacob Wolfson, Regina Soufli, Eberhard Spiller, Eric Gullikson, David McKenzie, David Windt, Leon Golub, William Podgorski, Paola Testa, and Mark Weber.
\newblock Initial calibration of the atmospheric imaging assembly ({AIA}) on the solar dynamics observatory ({SDO}).
\newblock \emph{Solar Physics}, 275\penalty0 (1):\penalty0 41--66, January 2012.

\bibitem[{Bonnet} et~al.(1978){Bonnet}, {Lemaire}, {Vial}, {Artzner}, {Gouttebroze}, {Jouchoux}, {Leibacher}, {Skumanich}, and {Vidal-Madjar}]{Bonnet1978LPSP_OSO8}
R.~M. {Bonnet}, P.~{Lemaire}, J.~C. {Vial}, G.~{Artzner}, P.~{Gouttebroze}, A.~{Jouchoux}, J.~W. {Leibacher}, A.~{Skumanich}, and A.~{Vidal-Madjar}.
\newblock {The LPSP instrument on OSO 8. II. In-flight performance and preliminary results.}
\newblock \emph{\apj}, 221:\penalty0 1032--1053, May 1978.
\newblock \doi{10.1086/156109}.

\bibitem[{Canfield} and {van Hoosier}(1980)]{CanfieldvanHoosier1980}
R.~C. {Canfield} and M.~E. {van Hoosier}.
\newblock {Observed L{\ensuremath{\alpha}} profiles for two solar flares: 14{\ensuremath{:}}12 UT 15 June, 1973 and 23{\ensuremath{:}}16 UT 21 January, 1974}.
\newblock \emph{\solphys}, 67\penalty0 (2):\penalty0 339--350, August 1980.
\newblock \doi{10.1007/BF00149811}.

\bibitem[{Chaffin} et~al.(2015){Chaffin}, {Chaufray}, {Deighan}, {Schneider}, {McClintock}, {Stewart}, {Thiemann}, {Clarke}, {Holsclaw}, {Jain}, {Crismani}, {Stiepen}, {Montmessin}, {Eparvier}, {Chamberlain}, and {Jakosky}]{Chaffin2015MAVEN-FISM-MApplication}
M.~S. {Chaffin}, J.~Y. {Chaufray}, J.~{Deighan}, N.~M. {Schneider}, W.~E. {McClintock}, A.~I.~F. {Stewart}, E.~{Thiemann}, J.~T. {Clarke}, G.~M. {Holsclaw}, S.~K. {Jain}, M.~M.~J. {Crismani}, A.~{Stiepen}, F.~{Montmessin}, F.~G. {Eparvier}, P.~C. {Chamberlain}, and B.~M. {Jakosky}.
\newblock {Three-dimensional structure in the Mars H corona revealed by IUVS on MAVEN}.
\newblock \emph{\grl}, 42\penalty0 (21):\penalty0 9001--9008, November 2015.
\newblock \doi{10.1002/2015GL065287}.

\bibitem[Chakraborty et~al.(2021)Chakraborty, Ruohoniemi, Baker, Fiori, Bailey, and Zawdie]{Chakraborty2021IonosphericSluggishness}
S.~Chakraborty, J.~M. Ruohoniemi, J.~B.~H. Baker, R.~A.~D. Fiori, S.~M. Bailey, and K.~A. Zawdie.
\newblock Ionospheric sluggishness: A characteristic time-lag of the ionospheric response to solar flares.
\newblock \emph{Journal of Geophysical Research: Space Physics}, 126\penalty0 (4):\penalty0 e2020JA028813, 2021.
\newblock \doi{https://doi.org/10.1029/2020JA028813}.
\newblock URL \url{https://agupubs.onlinelibrary.wiley.com/doi/abs/10.1029/2020JA028813}.
\newblock e2020JA028813 2020JA028813.

\bibitem[Chamberlin et~al.(2020)Chamberlin, Eparvier, Knoer, Leise, Pankratz, Snow, Templeman, Thiemann, Woodraska, and Woods]{Chamberlin2020FISM}
P.~C. Chamberlin, F.~G. Eparvier, V.~Knoer, H.~Leise, A.~Pankratz, M.~Snow, B.~Templeman, E.~M.~B. Thiemann, D.~L. Woodraska, and T.~N. Woods.
\newblock The flare irradiance spectral model-version 2 (fism2).
\newblock \emph{Space Weather}, 18\penalty0 (12):\penalty0 e2020SW002588, 2020.
\newblock \doi{https://doi.org/10.1029/2020SW002588}.
\newblock URL \url{https://agupubs.onlinelibrary.wiley.com/doi/abs/10.1029/2020SW002588}.
\newblock e2020SW002588 10.1029/2020SW002588.

\bibitem[{Chamberlin} et~al.(2020){Chamberlin}, {Schmit}, {Daw}, {Polito}, {Gong}, and {Milligan}]{Chamberlin2020SNIFS}
P.~C. {Chamberlin}, D.~J. {Schmit}, A.~N. {Daw}, V.~{Polito}, Q.~{Gong}, and R.~O. {Milligan}.
\newblock {The Solar eruptioN Integral Field Spectrograph (SNIFS) Sounding Rocket}.
\newblock In \emph{AGU Fall Meeting Abstracts}, volume 2020, pages SH056--03, December 2020.

\bibitem[{Chamberlin} et~al.(2009){Chamberlin}, {Woods}, {Eparvier}, and {Jones}]{Chamberlin2009XRS}
Phillip~C. {Chamberlin}, Thomas~N. {Woods}, Francis~G. {Eparvier}, and Andrew~R. {Jones}.
\newblock {Next generation x-ray sensor (XRS) for the NOAA GOES-R satellite series}.
\newblock In Silvano {Fineschi} and Judy~A. {Fennelly}, editors, \emph{Solar Physics and Space Weather Instrumentation III}, volume 7438 of \emph{Society of Photo-Optical Instrumentation Engineers (SPIE) Conference Series}, page 743802, August 2009.
\newblock \doi{10.1117/12.826807}.

\bibitem[{Chaufray} et~al.(2015){Chaufray}, {Deighan}, {Chaffin}, {Schneider}, {McClintock}, {Stewart}, {Jain}, {Crismani}, {Stiepen}, {Holsclaw}, {Clarke}, {Montmessin}, {Eparvier}, {Thiemann}, {Chamberlin}, and {Jakosky}]{Chaufray2015MAVEN-FISM-MApplication}
J.~Y. {Chaufray}, J.~{Deighan}, M.~S. {Chaffin}, N.~M. {Schneider}, W.~E. {McClintock}, A.~I.~F. {Stewart}, S.~K. {Jain}, M.~{Crismani}, A.~{Stiepen}, G.~M. {Holsclaw}, J.~T. {Clarke}, F.~{Montmessin}, F.~G. {Eparvier}, E.~M.~B. {Thiemann}, P.~C. {Chamberlin}, and B.~M. {Jakosky}.
\newblock {Study of the Martian cold oxygen corona from the O I 130.4 nm by IUVS/MAVEN}.
\newblock \emph{\grl}, 42\penalty0 (21):\penalty0 9031--9039, November 2015.
\newblock \doi{10.1002/2015GL065341}.

\bibitem[{Chen} et~al.(2024){Chen}, {Feng}, {Zhang}, {Li}, {He}, {Song}, {Guo}, {Li}, {Huang}, {Li}, {Zhao}, {Xue}, {Li}, {Shi}, {Song}, {Lu}, {Ying}, {Wang}, {Dai}, {Wang}, {Mao}, {Wang}, {Wu}, {Ren}, {Sun}, {Yang}, {Xia}, {Zhang}, {Zhou}, {Tao}, {Liu}, {Yu}, {Li}, {Li}, {Zhang}, {Li}, {Tian}, {Zhou}, {Tian}, {Shan}, {Liu}, {Jing}, and {Gan}]{Chen2024LST}
Bo~{Chen}, Li~{Feng}, Guang {Zhang}, Hui {Li}, Lingping {He}, Kefei {Song}, Quanfeng {Guo}, Ying {Li}, Yu~{Huang}, Jingwei {Li}, Jie {Zhao}, Jianchao {Xue}, Gen {Li}, Guanglu {Shi}, Dechao {Song}, Lei {Lu}, Beili {Ying}, Haifeng {Wang}, Shuang {Dai}, Xiaodong {Wang}, Shilei {Mao}, Peng {Wang}, Kun {Wu}, Shuai {Ren}, Liang {Sun}, Xianwei {Yang}, Mingyi {Xia}, Xiaoxue {Zhang}, Peng {Zhou}, Chen {Tao}, Yang {Liu}, Sibo {Yu}, Xinkai {Li}, Shuting {Li}, Ping {Zhang}, Qiao {Li}, Zhengyuan {Tian}, Yue {Zhou}, Jun {Tian}, Jiahui {Shan}, Xiaofeng {Liu}, Zhichen {Jing}, and Weiqun {Gan}.
\newblock {Inflight Performance and Calibrations of the Lyman-alpha Solar Telescope on board the Advanced Space-based Solar Observatory}.
\newblock \emph{arXiv e-prints}, art. arXiv:2408.01937, August 2024.
\newblock \doi{10.48550/arXiv.2408.01937}.

\bibitem[{Curdt} et~al.(2001){Curdt}, {Brekke}, {Feldman}, {Wilhelm}, {Dwivedi}, {Sch{\"u}hle}, and {Lemaire}]{Curdt2001LyaBrightest}
W.~{Curdt}, P.~{Brekke}, U.~{Feldman}, K.~{Wilhelm}, B.~N. {Dwivedi}, U.~{Sch{\"u}hle}, and P.~{Lemaire}.
\newblock {The SUMER spectral atlas of solar-disk features}.
\newblock \emph{\aap}, 375:\penalty0 591--613, August 2001.
\newblock \doi{10.1051/0004-6361:20010364}.

\bibitem[{da Costa} et~al.(2009){da Costa}, {Fletcher}, {Labrosse}, and {Zuccarello}]{deCosta2009NTLyaCorr}
F.~Rubio {da Costa}, L.~{Fletcher}, N.~{Labrosse}, and F.~{Zuccarello}.
\newblock Observations of a solar flare and filament eruption in lyman $\alpha$\ and x-rays.
\newblock \emph{A\&A}, 507\penalty0 (2):\penalty0 1005--1014, 2009.
\newblock \doi{10.1051/0004-6361/200912651}.
\newblock URL \url{https://doi.org/10.1051/0004-6361/200912651}.

\bibitem[{Dominique} et~al.(2013){Dominique}, {Hochedez}, {Schmutz}, {Dammasch}, {Shapiro}, {Kretzschmar}, {Zhukov}, {Gillotay}, {Stockman}, and {BenMoussa}]{Dominique2013LYRAInstr}
M.~{Dominique}, J.~F. {Hochedez}, W.~{Schmutz}, I.~E. {Dammasch}, A.~I. {Shapiro}, M.~{Kretzschmar}, A.~N. {Zhukov}, D.~{Gillotay}, Y.~{Stockman}, and A.~{BenMoussa}.
\newblock {The LYRA Instrument Onboard PROBA2: Description and In-Flight Performance}.
\newblock \emph{\solphys}, 286\penalty0 (1):\penalty0 21--42, August 2013.
\newblock \doi{10.1007/s11207-013-0252-5}.

\bibitem[{Dominique} et~al.(2018){Dominique}, {Zhukov}, {Heinzel}, {Dammasch}, {Wauters}, {Dolla}, {Shestov}, {Kretzschmar}, {Machol}, {Lapenta}, and {Schmutz}]{Dominique2018NTLyaCorr}
Marie {Dominique}, Andrei~N. {Zhukov}, Petr {Heinzel}, Ingolf~E. {Dammasch}, Laurence {Wauters}, Laurent {Dolla}, Sergei {Shestov}, Matthieu {Kretzschmar}, Janet {Machol}, Giovanni {Lapenta}, and Werner {Schmutz}.
\newblock {First Detection of Solar Flare Emission in Mid-ultraviolet Balmer Continuum}.
\newblock \emph{\apjl}, 867\penalty0 (2):\penalty0 L24, November 2018.
\newblock \doi{10.3847/2041-8213/aaeace}.

\bibitem[Eparvier et~al.(2015)Eparvier, Chamberlin, Woods, and Thiemann]{Eparvier2015MavenEUVM}
Francis Eparvier, Phillip Chamberlin, Thomas Woods, and Edward Thiemann.
\newblock The solar extreme ultraviolet monitor for maven.
\newblock \emph{Space Science Reviews}, 195, 09 2015.
\newblock \doi{10.1007/s11214-015-0195-2}.

\bibitem[{Eparvier} et~al.(2009){Eparvier}, {Crotser}, {Jones}, {McClintock}, {Snow}, and {Woods}]{Eparvier2009EUVS}
Francis~G. {Eparvier}, David {Crotser}, Andrew~R. {Jones}, William~E. {McClintock}, Martin {Snow}, and Thomas~N. {Woods}.
\newblock {The Extreme Ultraviolet Sensor (EUVS) for GOES-R}.
\newblock In Silvano {Fineschi} and Judy~A. {Fennelly}, editors, \emph{Solar Physics and Space Weather Instrumentation III}, volume 7438 of \emph{Society of Photo-Optical Instrumentation Engineers (SPIE) Conference Series}, page 743804, August 2009.
\newblock \doi{10.1117/12.826445}.

\bibitem[{Evans} et~al.(2010){Evans}, {Strickland}, {Woo}, {McMullin}, {Plunkett}, {Viereck}, {Hill}, {Woods}, and {Eparvier}]{Evans2010EUVS}
J.~S. {Evans}, D.~J. {Strickland}, W.~K. {Woo}, D.~R. {McMullin}, S.~P. {Plunkett}, R.~A. {Viereck}, S.~M. {Hill}, T.~N. {Woods}, and F.~G. {Eparvier}.
\newblock {Early Observations by the GOES-13 Solar Extreme Ultraviolet Sensor (EUVS)}.
\newblock \emph{\solphys}, 262\penalty0 (1):\penalty0 71--115, March 2010.
\newblock \doi{10.1007/s11207-009-9491-x}.

\bibitem[{Feng} et~al.(2019){Feng}, {Li}, {Chen}, {Li}, {Susino}, {Huang}, {Lu}, {Ying}, {Li}, {Xue}, {Yang}, {Hong}, {Li}, {Zhao}, {Gan}, and {Zhang}]{Feng2019LST}
Li~{Feng}, Hui {Li}, Bo~{Chen}, Ying {Li}, Roberto {Susino}, Yu~{Huang}, Lei {Lu}, Bei-Li {Ying}, Jing-Wei {Li}, Jian-Chao {Xue}, Yu-Tong {Yang}, Jie {Hong}, Jian-Ping {Li}, Jie {Zhao}, Wei-Qun {Gan}, and Yan {Zhang}.
\newblock {The Lyman-alpha Solar Telescope (LST) for the ASO-S mission - III. data and potential diagnostics}.
\newblock \emph{Research in Astronomy and Astrophysics}, 19\penalty0 (11):\penalty0 162, November 2019.
\newblock \doi{10.1088/1674-4527/19/11/162}.

\bibitem[{Gan} et~al.(2023){Gan}, {Zhu}, {Deng}, {Zhang}, {Chen}, {Huang}, {Deng}, {Wu}, {Zhang}, {Li}, {Su}, {Su}, {Feng}, {Wu}, {Cui}, {Wang}, {Chang}, {Yin}, {Xiong}, {Chen}, {Yang}, {Li}, {Lin}, {Hou}, {Bai}, {Chen}, {Zhang}, {Hu}, {Liang}, {Wang}, {Song}, {Guo}, {He}, {Zhang}, {Wang}, {Bao}, {Cao}, {Bai}, {Chen}, {He}, {Li}, {Zhang}, {Liao}, {Jiang}, {Li}, {Su}, {Lei}, {Chen}, {Li}, {Zhao}, {Li}, {Ge}, {Zou}, {Hu}, {Su}, {Ji}, {Gu}, {Zheng}, {Xu}, and {Wang}]{Gan2023ASOS}
Weiqun {Gan}, Cheng {Zhu}, Yuanyong {Deng}, Zhe {Zhang}, Bo~{Chen}, Yu~{Huang}, Lei {Deng}, Haiyan {Wu}, Haiying {Zhang}, Hui {Li}, Yang {Su}, Jiangtao {Su}, Li~{Feng}, Jian {Wu}, Jijun {Cui}, Chi {Wang}, Jin {Chang}, Zengshan {Yin}, Weiming {Xiong}, Bin {Chen}, Jianfeng {Yang}, Fu~{Li}, Jiaben {Lin}, Junfeng {Hou}, Xianyong {Bai}, Dengyi {Chen}, Yan {Zhang}, Yiming {Hu}, Yaoming {Liang}, Jianping {Wang}, Kefei {Song}, Quanfeng {Guo}, Lingping {He}, Guang {Zhang}, Peng {Wang}, Haicao {Bao}, Caixia {Cao}, Yanping {Bai}, Binglong {Chen}, Tao {He}, Xinyu {Li}, Ye~{Zhang}, Xing {Liao}, Hu~{Jiang}, Youping {Li}, Yingna {Su}, Shijun {Lei}, Wei {Chen}, Ying {Li}, Jie {Zhao}, Jingwei {Li}, Yunyi {Ge}, Ziming {Zou}, Tai {Hu}, Miao {Su}, Haidong {Ji}, Mei {Gu}, Yonghuang {Zheng}, Dezhen {Xu}, and Xing {Wang}.
\newblock {The Advanced Space-Based Solar Observatory (ASO-S)}.
\newblock \emph{\solphys}, 298\penalty0 (5):\penalty0 68, May 2023.
\newblock \doi{10.1007/s11207-023-02166-x}.

\bibitem[{Greatorex} et~al.(2023){Greatorex}, {Milligan}, and {Chamberlin}]{Greatorex2023EquivMagFlares}
Harry~J. {Greatorex}, Ryan~O. {Milligan}, and Phillip~C. {Chamberlin}.
\newblock {Observational Analysis of Ly{\ensuremath{\alpha}} Emission in Equivalent-magnitude Solar Flares}.
\newblock \emph{\apj}, 954\penalty0 (2):\penalty0 120, September 2023.
\newblock \doi{10.3847/1538-4357/acea7f}.

\bibitem[Hanser and Sellers(1996)]{Hanser1996GOES_XRS}
Frederick~A. Hanser and Francis~Bach Sellers.
\newblock {Design and calibration of the GOES-8 solar x-ray sensor: the XRS}.
\newblock In Edward~R. Washwell, editor, \emph{GOES-8 and Beyond}, volume 2812, pages 344 -- 352. International Society for Optics and Photonics, SPIE, 1996.
\newblock \doi{10.1117/12.254082}.
\newblock URL \url{https://doi.org/10.1117/12.254082}.

\bibitem[{Harra} et~al.(2022){Harra}, {Alberti}, {Berghmans}, {Chamberlin}, {Dominique}, {Eparvier}, {Gissot}, {Hara}, {Haberreiter}, {Hori}, {Jin}, {Kretzschmar}, {Imada}, {Kawate}, {Koller}, {Krucker}, {Langer}, {Meier}, {Milligan}, {Miyoshi}, {Nozomu}, {Nishitani}, {Pfiffner}, {Rozanov}, {Shimizu}, {Thiemann}, {Tye}, {Watanabe}, {Woods}, {Leng Yeo}, {Ieda}, and {Barczynski}]{Harra2022SoSPIMPreceedings}
Louise~K. {Harra}, Andrea {Alberti}, David {Berghmans}, Phillip {Chamberlin}, Marie {Dominique}, Francis~G. {Eparvier}, Samuel {Gissot}, Hirohisa {Hara}, Margit {Haberreiter}, Tomoaki {Hori}, Hidekatsu {Jin}, Matthieu {Kretzschmar}, Shinsuke {Imada}, Tomoko {Kawate}, Silvio {Koller}, Samuel {Krucker}, Patrick {Langer}, Leandro {Meier}, Ryan {Milligan}, Yoshizumi {Miyoshi}, {Nozomu}, {Nishitani}, Dany {Pfiffner}, Eugene {Rozanov}, Toshifumi {Shimizu}, Edward {Thiemann}, Daniel {Tye}, Kyoko {Watanabe}, Thomas {Woods}, Kok {Leng Yeo}, Akimasa {Ieda}, and Krzysztof {Barczynski}.
\newblock {A spectral solar irradiance monitor (SoSpIM) on the JAXA Solar-C (EUVST) space mission}.
\newblock In \emph{44th COSPAR Scientific Assembly. Held 16-24 July}, volume~44, page 834, July 2022.

\bibitem[Hayes et~al.(2021)Hayes, O’Hara, Murray, and Gallagher]{Hayes2021Ionosphere}
Laura~A. Hayes, Oscar S.~D. O’Hara, Sophie~A. Murray, and Peter~T. Gallagher.
\newblock Solar flare effects on the earth’s lower ionosphere.
\newblock \emph{Solar Physics}, 296\penalty0 (11), Nov 2021.
\newblock ISSN 1573-093X.
\newblock \doi{10.1007/s11207-021-01898-y}.
\newblock URL \url{http://dx.doi.org/10.1007/s11207-021-01898-y}.

\bibitem[{Hochedez} et~al.(2006){Hochedez}, {Schmutz}, {Stockman}, {Sch{\"u}hle}, {Benmoussa}, {Koller}, {Haenen}, {Berghmans}, {Defise}, {Halain}, {Theissen}, {Delouille}, {Slemzin}, {Gillotay}, {Fussen}, {Dominique}, {Vanhellemont}, {McMullin}, {Kretzschmar}, {Mitrofanov}, {Nicula}, {Wauters}, {Roth}, {Rozanov}, {R{\"u}edi}, {Wehrli}, {Soltani}, {Amano}, {van der Linden}, {Zhukov}, {Clette}, {Koizumi}, {Mortet}, {Remes}, {Petersen}, {Nesl{\'a}dek}, {D'Olieslaeger}, {Roggen}, and {Rochus}]{Hochedez2006PROBA2Lyra}
J.~F. {Hochedez}, W.~{Schmutz}, Y.~{Stockman}, U.~{Sch{\"u}hle}, A.~{Benmoussa}, S.~{Koller}, K.~{Haenen}, D.~{Berghmans}, J.~M. {Defise}, J.~P. {Halain}, A.~{Theissen}, V.~{Delouille}, V.~{Slemzin}, D.~{Gillotay}, D.~{Fussen}, M.~{Dominique}, F.~{Vanhellemont}, D.~{McMullin}, M.~{Kretzschmar}, A.~{Mitrofanov}, B.~{Nicula}, L.~{Wauters}, H.~{Roth}, E.~{Rozanov}, I.~{R{\"u}edi}, C.~{Wehrli}, A.~{Soltani}, H.~{Amano}, R.~{van der Linden}, A.~{Zhukov}, F.~{Clette}, S.~{Koizumi}, V.~{Mortet}, Z.~{Remes}, R.~{Petersen}, M.~{Nesl{\'a}dek}, M.~{D'Olieslaeger}, J.~{Roggen}, and P.~{Rochus}.
\newblock {LYRA, a solar UV radiometer on Proba2}.
\newblock \emph{Advances in Space Research}, 37\penalty0 (2):\penalty0 303--312, January 2006.
\newblock \doi{10.1016/j.asr.2005.10.041}.

\bibitem[{Howard} et~al.(2008){Howard}, {Moses}, {Vourlidas}, {Newmark}, {Socker}, {Plunkett}, {Korendyke}, {Cook}, {Hurley}, {Davila}, {Thompson}, {St Cyr}, {Mentzell}, {Mehalick}, {Lemen}, {Wuelser}, {Duncan}, {Tarbell}, {Wolfson}, {Moore}, {Harrison}, {Waltham}, {Lang}, {Davis}, {Eyles}, {Mapson-Menard}, {Simnett}, {Halain}, {Defise}, {Mazy}, {Rochus}, {Mercier}, {Ravet}, {Delmotte}, {Auchere}, {Delaboudiniere}, {Bothmer}, {Deutsch}, {Wang}, {Rich}, {Cooper}, {Stephens}, {Maahs}, {Baugh}, {McMullin}, and {Carter}]{Howard2008EUVI}
R.~A. {Howard}, J.~D. {Moses}, A.~{Vourlidas}, J.~S. {Newmark}, D.~G. {Socker}, S.~P. {Plunkett}, C.~M. {Korendyke}, J.~W. {Cook}, A.~{Hurley}, J.~M. {Davila}, W.~T. {Thompson}, O.~C. {St Cyr}, E.~{Mentzell}, K.~{Mehalick}, J.~R. {Lemen}, J.~P. {Wuelser}, D.~W. {Duncan}, T.~D. {Tarbell}, C.~J. {Wolfson}, A.~{Moore}, R.~A. {Harrison}, N.~R. {Waltham}, J.~{Lang}, C.~J. {Davis}, C.~J. {Eyles}, H.~{Mapson-Menard}, G.~M. {Simnett}, J.~P. {Halain}, J.~M. {Defise}, E.~{Mazy}, P.~{Rochus}, R.~{Mercier}, M.~F. {Ravet}, F.~{Delmotte}, F.~{Auchere}, J.~P. {Delaboudiniere}, V.~{Bothmer}, W.~{Deutsch}, D.~{Wang}, N.~{Rich}, S.~{Cooper}, V.~{Stephens}, G.~{Maahs}, R.~{Baugh}, D.~{McMullin}, and T.~{Carter}.
\newblock {Sun Earth Connection Coronal and Heliospheric Investigation (SECCHI)}.
\newblock \emph{\ssr}, 136\penalty0 (1-4):\penalty0 67--115, April 2008.
\newblock \doi{10.1007/s11214-008-9341-4}.

\bibitem[{Jain} et~al.(2015){Jain}, {Stewart}, {Schneider}, {Deighan}, {Stiepen}, {Evans}, {Stevens}, {Chaffin}, {Crismani}, {McClintock}, {Clarke}, {Holsclaw}, {Lo}, {Lef{\`e}vre}, {Montmessin}, {Thiemann}, {Eparvier}, and {Jakosky}]{Jain2015MAVEN-FISM-MApplication}
S.~K. {Jain}, A.~I.~F. {Stewart}, N.~M. {Schneider}, J.~{Deighan}, A.~{Stiepen}, J.~S. {Evans}, M.~H. {Stevens}, M.~S. {Chaffin}, M.~{Crismani}, W.~E. {McClintock}, J.~T. {Clarke}, G.~M. {Holsclaw}, D.~Y. {Lo}, F.~{Lef{\`e}vre}, F.~{Montmessin}, E.~M.~B. {Thiemann}, F.~{Eparvier}, and B.~M. {Jakosky}.
\newblock {The structure and variability of Mars upper atmosphere as seen in MAVEN/IUVS dayglow observations}.
\newblock \emph{\grl}, 42\penalty0 (21):\penalty0 9023--9030, November 2015.
\newblock \doi{10.1002/2015GL065419}.

\bibitem[{Kaiser} et~al.(2008){Kaiser}, {Kucera}, {Davila}, {St. Cyr}, {Guhathakurta}, and {Christian}]{Kaiser2008Stereo}
M.~L. {Kaiser}, T.~A. {Kucera}, J.~M. {Davila}, O.~C. {St. Cyr}, M.~{Guhathakurta}, and E.~{Christian}.
\newblock {The STEREO Mission: An Introduction}.
\newblock \emph{\ssr}, 136\penalty0 (1-4):\penalty0 5--16, April 2008.
\newblock \doi{10.1007/s11214-007-9277-0}.

\bibitem[{Kretzschmar} et~al.(2013){Kretzschmar}, {Dominique}, and {Dammasch}]{Kretzshmar2013LyraFLare}
M.~{Kretzschmar}, M.~{Dominique}, and I.~E. {Dammasch}.
\newblock {Sun-as-a-Star Observation of Flares in Lyman {\ensuremath{\alpha}} by the PROBA2/LYRA Radiometer}.
\newblock \emph{\solphys}, 286\penalty0 (1):\penalty0 221--239, August 2013.
\newblock \doi{10.1007/s11207-012-0175-6}.

\bibitem[{Li} et~al.(2024){Li}, {Hong}, {Hou}, and {Su}]{Dong2024LSTQPP}
Dong {Li}, Zhenxiang {Hong}, Zhenyong {Hou}, and Yang {Su}.
\newblock {Localizing Quasiperiodic Pulsations in Hard X-Ray, Microwave, and Ly{\ensuremath{\alpha}} Emissions of an X6.4 Flare}.
\newblock \emph{\apj}, 970\penalty0 (1):\penalty0 77, July 2024.
\newblock \doi{10.3847/1538-4357/ad566c}.

\bibitem[{Li} et~al.(2019){Li}, {Chen}, {Feng}, {Li}, {Huang}, {Li}, {Lu}, {Xue}, {Ying}, {Zhao}, {Yang}, {Gan}, {Fang}, {Song}, {Wang}, {Guo}, {He}, {Zhu}, {Zhu}, {Deng}, {Bao}, {Cao}, and {Yang}]{Li2019LST}
Hui {Li}, Bo~{Chen}, Li~{Feng}, Ying {Li}, Yu~{Huang}, Jing-Wei {Li}, Lei {Lu}, Jian-Chao {Xue}, Bei-Li {Ying}, Jie {Zhao}, Yu-Tong {Yang}, Wei-Qun {Gan}, Cheng {Fang}, Ke-Fei {Song}, Hong {Wang}, Quan-Feng {Guo}, Ling-Ping {He}, Bo~{Zhu}, Cheng {Zhu}, Lei {Deng}, Hai-Chao {Bao}, Cai-Xia {Cao}, and Zhong-Guang {Yang}.
\newblock {The Lyman-alpha Solar Telescope (LST) for the ASO-S mission {\textemdash} I. Scientific objectives and overview}.
\newblock \emph{Research in Astronomy and Astrophysics}, 19\penalty0 (11):\penalty0 158, November 2019.
\newblock \doi{10.1088/1674-4527/19/11/158}.

\bibitem[Li et~al.(2022)Li, Li, Song, Battaglia, Xiao, Krucker, Schühle, Li, Gan, and Ding]{Li2022C14SolO}
Y.~Li, Qiao Li, De-Chao Song, Andrea~Francesco Battaglia, Hualin Xiao, Säm Krucker, Udo Schühle, Hui Li, Weiqun Gan, and M.~D. Ding.
\newblock The ly{\ensuremath{\alpha}} emission in a c1.4 solar flare observed by the extreme ultraviolet imager aboard solar orbiter.
\newblock \emph{The Astrophysical Journal}, 936\penalty0 (2):\penalty0 142, sep 2022.
\newblock \doi{10.3847/1538-4357/ac897c}.
\newblock URL \url{https://dx.doi.org/10.3847/1538-4357/ac897c}.

\bibitem[{Lin} et~al.(2002){Lin}, {Dennis}, {Hurford}, {Smith}, {Zehnder}, {Harvey}, {Curtis}, {Pankow}, {Turin}, {Bester}, {Csillaghy}, {Lewis}, {Madden}, {van Beek}, {Appleby}, {Raudorf}, {McTiernan}, {Ramaty}, {Schmahl}, {Schwartz}, {Krucker}, {Abiad}, {Quinn}, {Berg}, {Hashii}, {Sterling}, {Jackson}, {Pratt}, {Campbell}, {Malone}, {Landis}, {Barrington-Leigh}, {Slassi-Sennou}, {Cork}, {Clark}, {Amato}, {Orwig}, {Boyle}, {Banks}, {Shirey}, {Tolbert}, {Zarro}, {Snow}, {Thomsen}, {Henneck}, {McHedlishvili}, {Ming}, {Fivian}, {Jordan}, {Wanner}, {Crubb}, {Preble}, {Matranga}, {Benz}, {Hudson}, {Canfield}, {Holman}, {Crannell}, {Kosugi}, {Emslie}, {Vilmer}, {Brown}, {Johns-Krull}, {Aschwanden}, {Metcalf}, and {Conway}]{Lin2002RHESSI}
R.~P. {Lin}, B.~R. {Dennis}, G.~J. {Hurford}, D.~M. {Smith}, A.~{Zehnder}, P.~R. {Harvey}, D.~W. {Curtis}, D.~{Pankow}, P.~{Turin}, M.~{Bester}, A.~{Csillaghy}, M.~{Lewis}, N.~{Madden}, H.~F. {van Beek}, M.~{Appleby}, T.~{Raudorf}, J.~{McTiernan}, R.~{Ramaty}, E.~{Schmahl}, R.~{Schwartz}, S.~{Krucker}, R.~{Abiad}, T.~{Quinn}, P.~{Berg}, M.~{Hashii}, R.~{Sterling}, R.~{Jackson}, R.~{Pratt}, R.~D. {Campbell}, D.~{Malone}, D.~{Landis}, C.~P. {Barrington-Leigh}, S.~{Slassi-Sennou}, C.~{Cork}, D.~{Clark}, D.~{Amato}, L.~{Orwig}, R.~{Boyle}, I.~S. {Banks}, K.~{Shirey}, A.~K. {Tolbert}, D.~{Zarro}, F.~{Snow}, K.~{Thomsen}, R.~{Henneck}, A.~{McHedlishvili}, P.~{Ming}, M.~{Fivian}, John {Jordan}, Richard {Wanner}, Jerry {Crubb}, J.~{Preble}, M.~{Matranga}, A.~{Benz}, H.~{Hudson}, R.~C. {Canfield}, G.~D. {Holman}, C.~{Crannell}, T.~{Kosugi}, A.~G. {Emslie}, N.~{Vilmer}, J.~C. {Brown}, C.~{Johns-Krull}, M.~{Aschwanden}, T.~{Metcalf}, and A.~{Conway}.
\newblock {The Reuven Ramaty High-Energy Solar Spectroscopic Imager (RHESSI)}.
\newblock \emph{\solphys}, 210\penalty0 (1):\penalty0 3--32, November 2002.
\newblock \doi{10.1023/A:1022428818870}.

\bibitem[Lollo et~al.(2012)Lollo, Withers, Fallows, Girazian, Matta, and Chamberlin]{Lollo2012FISMApplicationMars}
Anthony Lollo, Paul Withers, Kathryn Fallows, Zachary Girazian, Majd Matta, and P.~C. Chamberlin.
\newblock Numerical simulations of the ionosphere of mars during a solar flare.
\newblock \emph{Journal of Geophysical Research: Space Physics}, 117\penalty0 (A5), 2012.
\newblock \doi{https://doi.org/10.1029/2011JA017399}.
\newblock URL \url{https://agupubs.onlinelibrary.wiley.com/doi/abs/10.1029/2011JA017399}.

\bibitem[{McClintock} et~al.(2005){McClintock}, {Rottman}, and {Woods}]{McClintock2005SOLSTICE}
William~E. {McClintock}, Gary~J. {Rottman}, and Thomas~N. {Woods}.
\newblock {Solar-Stellar Irradiance Comparison Experiment II (Solstice II): Instrument Concept and Design}.
\newblock \emph{\solphys}, 230\penalty0 (1-2):\penalty0 225--258, August 2005.
\newblock \doi{10.1007/s11207-005-7432-x}.

\bibitem[Milligan(2021)]{Milligan2021B&CClassFlares}
Ryan~O. Milligan.
\newblock Solar irradiance variability due to solar flares observed in lyman-alpha emission.
\newblock \emph{Solar Physics}, 296\penalty0 (3), mar 2021.
\newblock \doi{10.1007/s11207-021-01796-3}.
\newblock URL \url{https://doi.org/10.1007%2Fs11207-021-01796-3}.

\bibitem[{Milligan} and {Chamberlin}(2016)]{Milligan2016SDO_EVE_MEGSP}
Ryan~O. {Milligan} and Phillip~C. {Chamberlin}.
\newblock {Anomalous temporal behaviour of broadband Ly{\ensuremath{\alpha}} observations during solar flares from SDO/EVE}.
\newblock \emph{\aap}, 587:\penalty0 A123, March 2016.
\newblock \doi{10.1051/0004-6361/201526682}.

\bibitem[Milligan et~al.(2014)Milligan, Kerr, Dennis, Hudson, Fletcher, Allred, Chamberlin, Ireland, Mathioudakis, and Keenan]{Milligan2014EUVEnergy}
Ryan~O. Milligan, Graham~S. Kerr, Brian~R. Dennis, Hugh~S. Hudson, Lyndsay Fletcher, Joel~C. Allred, Phillip~C. Chamberlin, Jack Ireland, Mihalis Mathioudakis, and Francis~P. Keenan.
\newblock The radiated energy budget of chromospheric plasma in a major solar flare deduced from multi-wavelength observations.
\newblock \emph{\apj}, 793\penalty0 (2):\penalty0 70, Sep 2014.
\newblock ISSN 1538-4357.
\newblock \doi{10.1088/0004-637x/793/2/70}.
\newblock URL \url{http://dx.doi.org/10.1088/0004-637X/793/2/70}.

\bibitem[Milligan et~al.(2017)Milligan, Fleck, Ireland, Fletcher, and Dennis]{Milligan2017NTLyaCorr}
Ryan~O. Milligan, Bernhard Fleck, Jack Ireland, Lyndsay Fletcher, and Brian~R. Dennis.
\newblock Detection of three-minute oscillations in full-disk ly{\ensuremath{\alpha}} emission during a solar flare.
\newblock \emph{The Astrophysical Journal}, 848\penalty0 (1):\penalty0 L8, oct 2017.
\newblock \doi{10.3847/2041-8213/aa8f3a}.
\newblock URL \url{https://doi.org/10.3847/2041-8213/aa8f3a}.

\bibitem[Milligan et~al.(2020)Milligan, Hudson, Chamberlin, Hannah, and Hayes]{Milligan2020MXFlares}
Ryan~O. Milligan, Hugh~S. Hudson, Phillip~C. Chamberlin, Iain~G. Hannah, and Laura~A. Hayes.
\newblock Lyman‐alpha variability during solar flares over solar cycle 24 using goes‐15/euvs‐e.
\newblock \emph{Space Weather}, 18\penalty0 (7), Jul 2020.
\newblock ISSN 1542-7390.
\newblock \doi{10.1029/2019sw002331}.
\newblock URL \url{http://dx.doi.org/10.1029/2019SW002331}.

\bibitem[{M\"uller} et~al.(2020){M\"uller}, {St. Cyr, O. C.}, {Zouganelis, I.}, {Gilbert, H. R.}, {Marsden, R.}, {Nieves-Chinchilla, T.}, {Antonucci, E.}, {Auch\`ere, F.}, {Berghmans, D.}, {Horbury, T. S.}, {Howard, R. A.}, {Krucker, S.}, {Maksimovic, M.}, {Owen, C. J.}, {Rochus, P.}, {Rodriguez-Pacheco, J.}, {Romoli, M.}, {Solanki, S. K.}, {Bruno, R.}, {Carlsson, M.}, {Fludra, A.}, {Harra, L.}, {Hassler, D. M.}, {Livi, S.}, {Louarn, P.}, {Peter, H.}, {Sch\"uhle, U.}, {Teriaca, L.}, {del Toro Iniesta, J. C.}, {Wimmer-Schweingruber, R. F.}, {Marsch, E.}, {Velli, M.}, {De Groof, A.}, {Walsh, A.}, and {Williams, D.}]{Muller2020SolarOrbiter}
D.~{M\"uller}, {St. Cyr, O. C.}, {Zouganelis, I.}, {Gilbert, H. R.}, {Marsden, R.}, {Nieves-Chinchilla, T.}, {Antonucci, E.}, {Auch\`ere, F.}, {Berghmans, D.}, {Horbury, T. S.}, {Howard, R. A.}, {Krucker, S.}, {Maksimovic, M.}, {Owen, C. J.}, {Rochus, P.}, {Rodriguez-Pacheco, J.}, {Romoli, M.}, {Solanki, S. K.}, {Bruno, R.}, {Carlsson, M.}, {Fludra, A.}, {Harra, L.}, {Hassler, D. M.}, {Livi, S.}, {Louarn, P.}, {Peter, H.}, {Sch\"uhle, U.}, {Teriaca, L.}, {del Toro Iniesta, J. C.}, {Wimmer-Schweingruber, R. F.}, {Marsch, E.}, {Velli, M.}, {De Groof, A.}, {Walsh, A.}, and {Williams, D.}
\newblock The solar orbiter mission - science overview.
\newblock \emph{A\&A}, 642:\penalty0 A1, 2020.
\newblock \doi{10.1051/0004-6361/202038467}.
\newblock URL \url{https://doi.org/10.1051/0004-6361/202038467}.

\bibitem[{Neupert}(1968)]{Neupert1968neuperteffect}
Werner~M. {Neupert}.
\newblock {Comparison of Solar X-Ray Line Emission with Microwave Emission during Flares}.
\newblock \emph{\apjl}, 153:\penalty0 L59, July 1968.
\newblock \doi{10.1086/180220}.

\bibitem[{Nusinov} et~al.(2006){Nusinov}, {Kazachevskaya}, {Kuznetsov}, {Myagkova}, and {Yushkov}]{Nusinov2006LyaNTCorr}
A.~A. {Nusinov}, T.~V. {Kazachevskaya}, S.~N. {Kuznetsov}, I.~N. {Myagkova}, and B.~Yu. {Yushkov}.
\newblock {Ultraviolet, hard X-ray, and gamma-ray emission of solar flares recorded by VUSS-L and SONG instruments in 2001 2003}.
\newblock \emph{Solar System Research}, 40\penalty0 (4):\penalty0 282--285, July 2006.
\newblock \doi{10.1134/S0038094606040034}.

\bibitem[{Pesnell} et~al.(2012){Pesnell}, {Thompson}, and {Chamberlin}]{Pesnell2012SDOMain}
W.~Dean {Pesnell}, B.~J. {Thompson}, and P.~C. {Chamberlin}.
\newblock {The Solar Dynamics Observatory (SDO)}.
\newblock \emph{\solphys}, 275\penalty0 (1-2):\penalty0 3--15, January 2012.
\newblock \doi{10.1007/s11207-011-9841-3}.

\bibitem[Qian et~al.(2010)Qian, Burns, Chamberlin, and Solomon]{Qian2010FISMApplication}
Liying Qian, Alan~G. Burns, Phillip~C. Chamberlin, and Stanley~C. Solomon.
\newblock Flare location on the solar disk: Modeling the thermosphere and ionosphere response.
\newblock \emph{Journal of Geophysical Research: Space Physics}, 115\penalty0 (A9), 2010.
\newblock \doi{https://doi.org/10.1029/2009JA015225}.
\newblock URL \url{https://agupubs.onlinelibrary.wiley.com/doi/abs/10.1029/2009JA015225}.

\bibitem[Qian et~al.(2011)Qian, Burns, Chamberlin, and Solomon]{Qian2011FISMApplication}
Liying Qian, Alan~G. Burns, Phillip~C. Chamberlin, and Stanley~C. Solomon.
\newblock Variability of thermosphere and ionosphere responses to solar flares.
\newblock \emph{Journal of Geophysical Research: Space Physics}, 116\penalty0 (A10), 2011.
\newblock \doi{https://doi.org/10.1029/2011JA016777}.
\newblock URL \url{https://agupubs.onlinelibrary.wiley.com/doi/abs/10.1029/2011JA016777}.

\bibitem[Qian et~al.(2012)Qian, Burns, Solomon, and Chamberlin]{Qian2012FISMApplication}
Liying Qian, Alan~G. Burns, Stanley~C. Solomon, and Phillip~C. Chamberlin.
\newblock Solar flare impacts on ionospheric electrodyamics.
\newblock \emph{Geophysical Research Letters}, 39\penalty0 (6), 2012.
\newblock \doi{https://doi.org/10.1029/2012GL051102}.
\newblock URL \url{https://agupubs.onlinelibrary.wiley.com/doi/abs/10.1029/2012GL051102}.

\bibitem[Raulin et~al.(2013)Raulin, Trottet, Kretzschmar, Macotela, Pacini, Bertoni, and Dammasch]{Raulin2013LyaIonosphere}
Jean-Pierre Raulin, Gérard Trottet, Matthieu Kretzschmar, Edith~L. Macotela, Alessandra Pacini, Fernando C.~P. Bertoni, and Ingolf~E. Dammasch.
\newblock Response of the low ionosphere to x-ray and lyman-$\alpha$ solar flare emissions.
\newblock \emph{Journal of Geophysical Research: Space Physics}, 118\penalty0 (1):\penalty0 570--575, 2013.
\newblock \doi{https://doi.org/10.1029/2012JA017916}.
\newblock URL \url{https://agupubs.onlinelibrary.wiley.com/doi/abs/10.1029/2012JA017916}.

\bibitem[{Rochus} et~al.(2020){Rochus}, {Auch\`ere, F.}, {Berghmans, D.}, {Harra, L.}, {Schmutz, W.}, {Sch\"uhle, U.}, {Addison, P.}, {Appourchaux, T.}, {Aznar Cuadrado, R.}, {Baker, D.}, {Barbay, J.}, {Bates, D.}, {BenMoussa, A.}, {Bergmann, M.}, {Beurthe, C.}, {Borgo, B.}, {Bonte, K.}, {Bouzit, M.}, {Bradley, L.}, {B\"uchel, V.}, {Buchlin, E.}, {B\"uchner, J.}, {Cab\'e, F.}, {Cadiergues, L.}, {Chaigneau, M.}, {Chares, B.}, {Choque Cortez, C.}, {Coker, P.}, {Condamin, M.}, {Coumar, S.}, {Curdt, W.}, {Cutler, J.}, {Davies, D.}, {Davison, G.}, {Defise, J.-M.}, {Del Zanna, G.}, {Delmotte, F.}, {Delouille, V.}, {Dolla, L.}, {Dumesnil, C.}, {D\"urig, F.}, {Enge, R.}, {Fran\c{c}ois, S.}, {Fourmond, J.-J.}, {Gillis, J.-M.}, {Giordanengo, B.}, {Gissot, S.}, {Green, L. M.}, {Guerreiro, N.}, {Guilbaud, A.}, {Gyo, M.}, {Haberreiter, M.}, {Hafiz, A.}, {Hailey, M.}, {Halain, J.-P.}, {Hansotte, J.}, {Hecquet, C.}, {Heerlein, K.}, {Hellin, M.-L.}, {Hemsley, S.}, {Hermans, A.}, {Hervier, V.}, {Hochedez, J.-F.}, {Houbrechts,
  Y.}, {Ihsan, K.}, {Jacques, L.}, {J\'er\^ome, A.}, {Jones, J.}, {Kahle, M.}, {Kennedy, T.}, {Klaproth, M.}, {Kolleck, M.}, {Koller, S.}, {Kotsialos, E.}, {Kraaikamp, E.}, {Langer, P.}, {Lawrenson, A.}, {Le Clech\'{}, J.-C.}, {Lenaerts, C.}, {Liebecq, S.}, {Linder, D.}, {Long, D. M.}, {Mampaey, B.}, {Markiewicz-Innes, D.}, {Marquet, B.}, {Marsch, E.}, {Matthews, S.}, {Mazy, E.}, {Mazzoli, A.}, {Meining, S.}, {Meltchakov, E.}, {Mercier, R.}, {Meyer, S.}, {Monecke, M.}, {Monfort, F.}, {Morinaud, G.}, {Moron, F.}, {Mountney, L.}, {M\"uller, R.}, {Nicula, B.}, {Parenti, S.}, {Peter, H.}, {Pfiffner, D.}, {Philippon, A.}, {Phillips, I.}, {Plesseria, J.-Y.}, {Pylyser, E.}, {Rabecki, F.}, {Ravet-Krill, M.-F.}, {Rebellato, J.}, {Renotte, E.}, {Rodriguez, L.}, {Roose, S.}, {Rosin, J.}, {Rossi, L.}, {Roth, P.}, {Rouesnel, F.}, {Roulliay, M.}, {Rousseau, A.}, {Ruane, K.}, {Scanlan, J.}, {Schlatter, P.}, {Seaton, D. B.}, {Silliman, K.}, {Smit, S.}, {Smith, P. J.}, {Solanki, S. K.}, {Spescha, M.}, {Spencer, A.}, {Stegen,
  K.}, {Stockman, Y.}, {Szwec, N.}, {Tamiatto, C.}, {Tandy, J.}, {Teriaca, L.}, {Theobald, C.}, {Tychon, I.}, {van Driel-Gesztelyi, L.}, {Verbeeck, C.}, {Vial, J.-C.}, {Werner, S.}, {West, M. J.}, {Westwood, D.}, {Wiegelmann, T.}, {Willis, G.}, {Winter, B.}, {Zerr, A.}, {Zhang, X.}, and {Zhukov, A. N.}]{Rochus2020EUI}
P.~{Rochus}, {Auch\`ere, F.}, {Berghmans, D.}, {Harra, L.}, {Schmutz, W.}, {Sch\"uhle, U.}, {Addison, P.}, {Appourchaux, T.}, {Aznar Cuadrado, R.}, {Baker, D.}, {Barbay, J.}, {Bates, D.}, {BenMoussa, A.}, {Bergmann, M.}, {Beurthe, C.}, {Borgo, B.}, {Bonte, K.}, {Bouzit, M.}, {Bradley, L.}, {B\"uchel, V.}, {Buchlin, E.}, {B\"uchner, J.}, {Cab\'e, F.}, {Cadiergues, L.}, {Chaigneau, M.}, {Chares, B.}, {Choque Cortez, C.}, {Coker, P.}, {Condamin, M.}, {Coumar, S.}, {Curdt, W.}, {Cutler, J.}, {Davies, D.}, {Davison, G.}, {Defise, J.-M.}, {Del Zanna, G.}, {Delmotte, F.}, {Delouille, V.}, {Dolla, L.}, {Dumesnil, C.}, {D\"urig, F.}, {Enge, R.}, {Fran\c{c}ois, S.}, {Fourmond, J.-J.}, {Gillis, J.-M.}, {Giordanengo, B.}, {Gissot, S.}, {Green, L. M.}, {Guerreiro, N.}, {Guilbaud, A.}, {Gyo, M.}, {Haberreiter, M.}, {Hafiz, A.}, {Hailey, M.}, {Halain, J.-P.}, {Hansotte, J.}, {Hecquet, C.}, {Heerlein, K.}, {Hellin, M.-L.}, {Hemsley, S.}, {Hermans, A.}, {Hervier, V.}, {Hochedez, J.-F.}, {Houbrechts, Y.}, {Ihsan, K.},
  {Jacques, L.}, {J\'er\^ome, A.}, {Jones, J.}, {Kahle, M.}, {Kennedy, T.}, {Klaproth, M.}, {Kolleck, M.}, {Koller, S.}, {Kotsialos, E.}, {Kraaikamp, E.}, {Langer, P.}, {Lawrenson, A.}, {Le Clech\'{}, J.-C.}, {Lenaerts, C.}, {Liebecq, S.}, {Linder, D.}, {Long, D. M.}, {Mampaey, B.}, {Markiewicz-Innes, D.}, {Marquet, B.}, {Marsch, E.}, {Matthews, S.}, {Mazy, E.}, {Mazzoli, A.}, {Meining, S.}, {Meltchakov, E.}, {Mercier, R.}, {Meyer, S.}, {Monecke, M.}, {Monfort, F.}, {Morinaud, G.}, {Moron, F.}, {Mountney, L.}, {M\"uller, R.}, {Nicula, B.}, {Parenti, S.}, {Peter, H.}, {Pfiffner, D.}, {Philippon, A.}, {Phillips, I.}, {Plesseria, J.-Y.}, {Pylyser, E.}, {Rabecki, F.}, {Ravet-Krill, M.-F.}, {Rebellato, J.}, {Renotte, E.}, {Rodriguez, L.}, {Roose, S.}, {Rosin, J.}, {Rossi, L.}, {Roth, P.}, {Rouesnel, F.}, {Roulliay, M.}, {Rousseau, A.}, {Ruane, K.}, {Scanlan, J.}, {Schlatter, P.}, {Seaton, D. B.}, {Silliman, K.}, {Smit, S.}, {Smith, P. J.}, {Solanki, S. K.}, {Spescha, M.}, {Spencer, A.}, {Stegen, K.}, {Stockman,
  Y.}, {Szwec, N.}, {Tamiatto, C.}, {Tandy, J.}, {Teriaca, L.}, {Theobald, C.}, {Tychon, I.}, {van Driel-Gesztelyi, L.}, {Verbeeck, C.}, {Vial, J.-C.}, {Werner, S.}, {West, M. J.}, {Westwood, D.}, {Wiegelmann, T.}, {Willis, G.}, {Winter, B.}, {Zerr, A.}, {Zhang, X.}, and {Zhukov, A. N.}
\newblock The solar orbiter eui instrument: The extreme ultraviolet imager.
\newblock \emph{A\&A}, 642:\penalty0 A8, 2020.
\newblock \doi{10.1051/0004-6361/201936663}.
\newblock URL \url{https://doi.org/10.1051/0004-6361/201936663}.

\bibitem[{Sakai} et~al.(2016){Sakai}, {Andersson}, {Cravens}, {Mitchell}, {Mazelle}, {Rahmati}, {Fowler}, {Bougher}, {Thiemann}, {Eparvier}, {Fontenla}, {Mahaffy}, {Connerney}, and {Jakosky}]{Sakai2016MAVEN-FISM-MApplication}
Shotaro {Sakai}, Laila {Andersson}, Thomas~E. {Cravens}, David~L. {Mitchell}, Christian {Mazelle}, Ali {Rahmati}, Christopher~M. {Fowler}, Stephen~W. {Bougher}, Edward M.~B. {Thiemann}, Francis~G. {Eparvier}, Juan~M. {Fontenla}, Paul~R. {Mahaffy}, John E.~P. {Connerney}, and Bruce~M. {Jakosky}.
\newblock {Electron energetics in the Martian dayside ionosphere: Model comparisons with MAVEN data}.
\newblock \emph{Journal of Geophysical Research (Space Physics)}, 121\penalty0 (7):\penalty0 7049--7066, July 2016.
\newblock \doi{10.1002/2016JA022782}.

\bibitem[{Santandrea} et~al.(2013){Santandrea}, {Gantois}, {Strauch}, {Teston}, {Tilmans}, {Baijot}, {Gerrits}, {De Groof}, {Schwehm}, and {Zender}]{Santandrea2013Proba2}
S.~{Santandrea}, K.~{Gantois}, K.~{Strauch}, F.~{Teston}, E.~{Tilmans}, C.~{Baijot}, D.~{Gerrits}, A.~{De Groof}, G.~{Schwehm}, and J.~{Zender}.
\newblock {PROBA2: Mission and Spacecraft Overview}.
\newblock \emph{\solphys}, 286\penalty0 (1):\penalty0 5--19, August 2013.
\newblock \doi{10.1007/s11207-013-0289-5}.

\bibitem[{Skumanich} et~al.(1978){Skumanich}, {Jouchoux}, {Castelli}, {Lemaire}, {Artzner}, {Gouttebroze}, {Vial}, and {Bonnet}]{Skumanich1978FlareLya}
A.~{Skumanich}, A.~{Jouchoux}, J.~{Castelli}, P.~{Lemaire}, G.~{Artzner}, P.~{Gouttebroze}, J.~C. {Vial}, and R.~M. {Bonnet}.
\newblock {OSO-8 Radio and X-ray observations of the 19 April 1977 flare.}
\newblock In \emph{Bulletin of the American Astronomical Society}, volume~10, page 441, March 1978.

\bibitem[{The SunPy Community} et~al.(2020){The SunPy Community}, Barnes, Bobra, Christe, Freij, Hayes, Ireland, Mumford, Perez-Suarez, Ryan, Shih, Chanda, Glogowski, Hewett, Hughitt, Hill, Hiware, Inglis, Kirk, Konge, Mason, Maloney, Murray, Panda, Park, Pereira, Reardon, Savage, Sipőcz, Stansby, Jain, Taylor, Yadav, Rajul, and Dang]{SunPy2020}
{The SunPy Community}, Will~T. Barnes, Monica~G. Bobra, Steven~D. Christe, Nabil Freij, Laura~A. Hayes, Jack Ireland, Stuart Mumford, David Perez-Suarez, Daniel~F. Ryan, Albert~Y. Shih, Prateek Chanda, Kolja Glogowski, Russell Hewett, V.~Keith Hughitt, Andrew Hill, Kaustubh Hiware, Andrew Inglis, Michael S.~F. Kirk, Sudarshan Konge, James~Paul Mason, Shane~Anthony Maloney, Sophie~A. Murray, Asish Panda, Jongyeob Park, Tiago M.~D. Pereira, Kevin Reardon, Sabrina Savage, Brigitta~M. Sipőcz, David Stansby, Yash Jain, Garrison Taylor, Tannmay Yadav, Rajul, and Trung~Kien Dang.
\newblock The sunpy project: Open source development and status of the version 1.0 core package.
\newblock \emph{The Astrophysical Journal}, 890:\penalty0 68--, 2020.
\newblock \doi{10.3847/1538-4357/ab4f7a}.
\newblock URL \url{https://iopscience.iop.org/article/10.3847/1538-4357/ab4f7a}.

\bibitem[Thiemann et~al.(2017)Thiemann, Chamberlin, Eparvier, Templeman, Woods, Bougher, and Jakosky]{Thiemann2017MavenEUVMModels}
Edward M.~B. Thiemann, Phillip~C. Chamberlin, Francis~G. Eparvier, Brian Templeman, Thomas~N. Woods, Stephen~W. Bougher, and Bruce~M. Jakosky.
\newblock The maven euvm model of solar spectral irradiance variability at mars: Algorithms and results.
\newblock \emph{Journal of Geophysical Research: Space Physics}, 122\penalty0 (3):\penalty0 2748--2767, 2017.
\newblock \doi{https://doi.org/10.1002/2016JA023512}.
\newblock URL \url{https://agupubs.onlinelibrary.wiley.com/doi/abs/10.1002/2016JA023512}.

\bibitem[{Tian} et~al.(2023){Tian}, {Feng}, {Lu}, {Xia}, {Su}, {Gan}, {Li}, and {Zhou}]{Tian2023ntlya}
Zheng-Yuan {Tian}, Li~{Feng}, Lei {Lu}, Fan-Xiaoyu {Xia}, Yang {Su}, Wei-Qun {Gan}, Hui {Li}, and Yue {Zhou}.
\newblock {Ly{\ensuremath{\alpha}} Emission Enhancement Associated with Soft X-Ray Microflares}.
\newblock \emph{Research in Astronomy and Astrophysics}, 23\penalty0 (6):\penalty0 065011, June 2023.
\newblock \doi{10.1088/1674-4527/accc75}.

\bibitem[{van Dokkum} et~al.(2012){van Dokkum}, {Bloom}, and {Tewes}]{vanDokkum2012LAcosmic}
Pieter~G. {van Dokkum}, J.~{Bloom}, and Malte {Tewes}.
\newblock {L.A.Cosmic: Laplacian Cosmic Ray Identification}.
\newblock Astrophysics Source Code Library, record ascl:1207.005, July 2012.

\bibitem[{Viereck} et~al.(2007){Viereck}, {Hanser}, {Wise}, {Guha}, {Jones}, {McMullin}, {Plunket}, {Strickland}, and {Evans}]{Viereck2007EUVS}
Rodney {Viereck}, Fred {Hanser}, John {Wise}, Soumyendu {Guha}, Andrew {Jones}, Don {McMullin}, Simon {Plunket}, Doug {Strickland}, and Scott {Evans}.
\newblock {Solar extreme ultraviolet irradiance observations from GOES: design characteristics and initial performance}.
\newblock In Silvano {Fineschi} and Rodney~A. {Viereck}, editors, \emph{Solar Physics and Space Weather Instrumentation II}, volume 6689 of \emph{Society of Photo-Optical Instrumentation Engineers (SPIE) Conference Series}, page 66890K, September 2007.
\newblock \doi{10.1117/12.734886}.

\bibitem[Watanabe(2014)]{Watanabe2014SolarC}
Tetsuya Watanabe.
\newblock {The Solar-C Mission}.
\newblock In Jacobus M.~Oschmann Jr., Mark Clampin, Giovanni~G. Fazio, and Howard~A. MacEwen, editors, \emph{Space Telescopes and Instrumentation 2014: Optical, Infrared, and Millimeter Wave}, volume 9143, page 91431O. International Society for Optics and Photonics, SPIE, 2014.
\newblock \doi{10.1117/12.2055366}.
\newblock URL \url{https://doi.org/10.1117/12.2055366}.

\bibitem[{Wauters} et~al.(2022){Wauters}, {Dominique}, {Milligan}, {Dammasch}, {Kretzschmar}, and {Machol}]{Wauters2022M67Lya}
L.~{Wauters}, M.~{Dominique}, R.~{Milligan}, I.~E. {Dammasch}, M.~{Kretzschmar}, and J.~{Machol}.
\newblock {Observation of a Flare and Filament Eruption in Lyman-{\ensuremath{\alpha}} on 8 September 2011 by the PRoject for OnBoard Autonomy/Large Yield Radiometer (PROBA2/LYRA)}.
\newblock \emph{\solphys}, 297\penalty0 (3):\penalty0 36, March 2022.
\newblock \doi{10.1007/s11207-022-01963-0}.

\bibitem[{Woods} et~al.(2012){Woods}, {Eparvier}, {Hock}, {Jones}, {Woodraska}, {Judge}, {Didkovsky}, {Lean}, {Mariska}, {Warren}, {McMullin}, {Chamberlin}, {Berthiaume}, {Bailey}, {Fuller-Rowell}, {Sojka}, {Tobiska}, and {Viereck}]{Woods2012EVE}
T.~N. {Woods}, F.~G. {Eparvier}, R.~{Hock}, A.~R. {Jones}, D.~{Woodraska}, D.~{Judge}, L.~{Didkovsky}, J.~{Lean}, J.~{Mariska}, H.~{Warren}, D.~{McMullin}, P.~{Chamberlin}, G.~{Berthiaume}, S.~{Bailey}, T.~{Fuller-Rowell}, J.~{Sojka}, W.~K. {Tobiska}, and R.~{Viereck}.
\newblock {Extreme Ultraviolet Variability Experiment (EVE) on the Solar Dynamics Observatory (SDO): Overview of Science Objectives, Instrument Design, Data Products, and Model Developments}.
\newblock \emph{\solphys}, 275\penalty0 (1-2):\penalty0 115--143, January 2012.
\newblock \doi{10.1007/s11207-009-9487-6}.

\bibitem[{Woods} et~al.(1995){Woods}, {Rottman}, {White}, {Fontenla}, and {Avrett}]{Woods1995CLV}
Thomas~N. {Woods}, Gary~J. {Rottman}, O.~R. {White}, Juan {Fontenla}, and E.~H. {Avrett}.
\newblock {The Solar LY alpha Line Profile}.
\newblock \emph{\apj}, 442:\penalty0 898, April 1995.
\newblock \doi{10.1086/175492}.

\bibitem[Woods et~al.(2006)Woods, Kopp, and Chamberlin]{Woods2006TSIVariation}
Thomas~N. Woods, Greg Kopp, and Phillip~C. Chamberlin.
\newblock Contributions of the solar ultraviolet irradiance to the total solar irradiance during large flares.
\newblock \emph{Journal of Geophysical Research: Space Physics}, 111\penalty0 (A10), 2006.
\newblock \doi{https://doi.org/10.1029/2005JA011507}.
\newblock URL \url{https://agupubs.onlinelibrary.wiley.com/doi/abs/10.1029/2005JA011507}.

\bibitem[{Woods} et~al.(2009){Woods}, {Chamberlin}, {Harder}, {Hock}, {Snow}, {Eparvier}, {Fontenla}, {McClintock}, and {Richard}]{Woods2009WHI}
Thomas~N. {Woods}, Phillip~C. {Chamberlin}, Jerald~W. {Harder}, Rachel~A. {Hock}, Martin {Snow}, Francis~G. {Eparvier}, Juan {Fontenla}, William~E. {McClintock}, and Erik~C. {Richard}.
\newblock {Solar Irradiance Reference Spectra (SIRS) for the 2008 Whole Heliosphere Interval (WHI)}.
\newblock \emph{\grl}, 36\penalty0 (1):\penalty0 L01101, January 2009.
\newblock \doi{10.1029/2008GL036373}.

\end{thebibliography}

%
%
%
%

\end{document}